\documentclass[final,5p,times,twocolumn]{elsarticle}

\apptocmd{\bibliography}{\raggedright\sloppy}{}{}
\usepackage{etoolbox}
\usepackage{dirtytalk}
\usepackage{comment}
\usepackage{import}
\usepackage{xurl}
\usepackage[hidelinks]{hyperref}
\usepackage{subcaption}
\usepackage{amssymb,amsthm,amsmath}
\usepackage{xcolor,paralist,fancyhdr}
\usepackage{xr}

\usepackage[english]{babel}

\journal{Future Generation Computer Systems}

\begin{document}


\title{A new approach to rating scale definition with quantum-inspired optimization}


\affiliation[affINFN]{organization={INFN Ferrara},
                      postcode={44122},
                      city={Ferrara},
                      state={Italy}}
\affiliation[affisp]{organization={Intesa Sanpaolo},
                      city={Torino},
                      state={Italy}}
\affiliation[afffis-ferr]{organization={Department of Physics and Earth Science, University of Ferrara},
                      postcode={44122},
                      city={Ferrara},
                      state={Italy}}
\affiliation[affsci]{organization={Department of Environment and Prevention Sciences, University of Ferrara},
                      postcode={44121},
                      city={Ferrara},
                      state={Italy}}

\author[affINFN]{Patrizio Spada\fnref{label1}}
\fntext[label1]{Present addresses: Department of Physics and Chemistry - Emilio Segrè,
University of Palermo, 90123, Palermo, Italy;
INAF - Arcetri Astrophysical Observatory,
50125,
Florence,
Italy}

\cortext[corauth]{Corresponding author: Laura Cappelli (laura.cappelli@fe.infn.it)}
\author[affINFN]{Laura Cappelli\corref{corauth}}
\author[affisp]{Francesca Cibrario}
\author[affisp]{Christian Mattia}
\author[affisp]{Daniele Magnaldi}
\author[affINFN,afffis-ferr]{Matteo Argenton}
\author[affINFN]{Enrico Calore}
\author[affINFN,affsci]{Sebastiano Fabio Schifano}
\author[affINFN]{Concezio Bozzi}
\author[affisp]{Davide Corbelletto}

\begin{frontmatter}

\begin{abstract}
In finance, assessing the creditworthiness of loan applicants requires lenders to cluster borrowers using rating scales. 
Financial institutions must define the scales in compliance with strict institutional constraints, resulting in solving a complex combinatorial constrained optimization problem. 
This contribution studies how to solve this problem using a Quadratic Unconstrained Binary Optimization (QUBO) model, a formulation suitable for quantum hardware. 
We validate this approach by testing the proposed formulation with classical heuristics. We then benchmark the results against a brute-force method to demonstrate consistent solution quality and highlight the framework’s suitability for more complex scenarios.
\end{abstract}


\end{frontmatter}


\section{Introduction}
\subsection{Rating scale definition}

Banking institutions aim to improve financial inclusion, credit access, and risk assessment efficiency. Therefore, they devise and adopt strategies to estimate the creditworthiness of their counterparts\textendash namely, loan borrowers\textendash by evaluating credit scores.
The credit scoring process ranks and clusters counterparts based on qualitative and quantitative characteristics using traditional statistics or machine learning methods, leveraging historical data.
Using these approaches to evaluate scores reduces human intervention and increases objectivity in the credit granting and monitoring process.

Using a database of counterparts, where each entity is defined by a set of attributes and a target variable indicating whether a default event has occurred, the creation of a credit scoring system involves three main stages:
\begin{enumerate}
    \item selecting a set of attributes: choosing the most informative subset of attributes that characterize the counterparts;
    \item computing the credit score: assigning to each counterpart a score representing its likelihood of default, or equivalently, the inverse of creditworthiness. At the end of this stage, counterparts can be ordered by score, with the most creditworthy at the top and the least creditworthy at the bottom;
    \item grouping counterparts into sets: counterparts are grouped into rating categories, called \textit{rating grades}.
\end{enumerate}
This manuscript addresses the third stage, the definition of the rating scale. A rating scale is defined as a partition of a sequence of $n$ counterparts ordered by score into $m$ nonempty sets, according to managerial and regulatory constraints, some of which are provided by European Union laws and guidance~\cite{gim,eba}. The number of ways to define grades without considering constraints is given by the binomial coefficient $\binom{n-1}{m-1}$.
For a new credit applicant for whom the score, grade, and likelihood of default are unknown, the credit scoring system can determine its score and grade.

\subsection{Rating scale definition and quantum computing}

The central research question that motivated the study presented in this article can be formulated as follows: can quantum computers offer an advantage for the definition of the rating scale? More specifically, this work investigates whether quantum-inspired paradigms can enhance existing credit rating assessment methodologies by improving computational efficiency or offering novel analytical perspectives.

An intuitive way to approach the problem of defining a rating scale is through clustering techniques. Among the quantum algorithms designed to solve these types of problems, quantum k-means is a natural candidate. Noisy Intermediate-Scale Quantum (NISQ)~\cite{Preskill} implementations of k-means clustering have been proposed in~\cite{kmeanclu,quex,qclu}. However, when applied to the present context, this algorithm shifts much of the overall complexity to constructing quantum circuits capable of encoding the relevant financial constraints. Due to the inherent intricacy of these constraints, we were therefore motivated to explore alternative methods.

In recent years, a mathematical formulation known as QUBO (Quadratic Unconstrained Binary Optimization)~\cite{Punnen} has been shown to encompass a wide variety of important combinatorial optimization (CO) problems~\cite{qubo,setprob,vertcolor,twosat}. 
For the problem at hand, this formulation represents an alternative approach to clustering due to its connection with CO problems and, specifically, the combinatorial nature of the rating scale definition. The QUBO formulation employs binary decision variables and a quadratic objective function that can be minimized or maximized without imposing constraints.
Institutional constraints on the rating scale definition are expressed as equations and inequalities. This makes the problem a discrete constrained optimization task. Such a formulation can be translated into an equivalent unconstrained problem by embedding the constraints directly into the objective function.

Although QUBO problems are NP-hard, meaning that they cannot be solved in feasible time for large instances, classical optimization algorithms, such as branch-and-bound~\cite{auth_method}, are designed to solve them more efficiently. In addition, quantum computing offers a novel perspective for tackling these problems due to the equivalence between the QUBO formulation and the Ising hamiltonian.
The feasibility of this approach has been explored across a variety of hardware architectures, including gate-based systems, quantum annealers, and Rydberg atom arrays.

The following sections present and validate a novel QUBO formulation for defining a rating scale. Specifically, Section~\ref{sec:related-works} reviews the literature on the application of quantum computing to credit scoring systems and provides an overview of quantum optimization. Section~\ref{section:formulation} details the financial constraints and explains the definition of the objective function. Section~\ref{sec_results} focuses on the formulation's implementation, describing the solvers and the results obtained. Additionally, this section highlights the advantages of the QUBO approach over a constrained combinatorial algorithm. Finally, Section~\ref{sec_conclusion} critically discusses the results and outlines potential directions for future work.
\section{Related works}
\label{sec:related-works}

\subsection{Credit Scoring}
Credit scoring has been extensively studied in the financial risk management literature, with a strong focus on statistical robustness, interpretability, and regulatory compliance. Classical approaches rely on statistical and machine learning techniques, for which comprehensive surveys can be found in the literature \cite{Abdou_Pointon_2011}. To identify quantum-based alternatives to conventional methodologies, we conducted a structured review of the existing literature, organizing our analysis around the key steps required to develop a robust credit scoring system, as detailed below. 
\begin{enumerate}
\item Selecting a set of attributes: this step aims to reduce the dimensionality of the data. Lenders and financial
institutions usually dispose of a large amount of data that can be used to assess creditworthiness. It
is therefore important to accurately select the most informative set of attributes to be used to
evaluate the counterparts. The objective is to simplify models without losing significant information, improve computational efficiency, and sometimes enhance model performance by mitigating overfitting. The
brute-force approach to this problem grows exponentially with the size of the
feature set, since each combination of attributes should be tested to find the best one. An extensive review of
feature selection traditional techniques can be found in \cite{Jia2022FeatureDR}.

More recently, quantum computing has been explored as a promising paradigm for feature selection. In particular, the authors of \cite{Paper2017OptimalFS} showed that the optimal balance between feature influence and
feature independence can be expressed by a QUBO problem.  Implementing the QUBO formulation on a quantum
annealer has the potential to outperform classical solvers.

Feature selection has also been addressed using digital quantum computing. For instance, the variational quantum feature selection (VarQFS)
algorithm introduced in \cite{Zoufal2023variationalquantum} relies on training a parameterized quantum circuit to identify informative feature subsets, optimized using Quantum Natural SPSA (QN-SPSA). When compared to traditional classical techniques VarQFS produces competitive feature subsets.

Finally, hybrid approaches have also been proposed. In \cite{Chen_Tso_He_2024}, a Quantum-Inspired Evolutionary Algorithm (QIEA) combines concepts from quantum computing and evolutionary computation, achieving more efficient solutions than conventional evolutionary or genetic algorithms.

\item
Computing the credit score: this step consists of training a model to predict the probability of default for a
given instance, leveraging historical borrower data. A Boolean label indicating default occurrence
can be employed to determine whether a default has occurred or not. The process of assigning a
credit score can be modeled as a binary classification problem, where the target variable is the
default status of borrowers.
Quantum computing techniques can be applied to the classification task. For credit scoring,
quantum classifiers can analyze complex patterns and relationships within the data that may
be challenging for classical classifiers to detect. Quantum neural networks (QNN) have been developed for this purpose. The integration of quantum
circuits with classical neural networks for enhancing credit scoring for small and medium sized
enterprises (SMEs) has been investigated in \cite{math12091391} achieving notable reductions in
training time for a similar predictive accuracy. A decrease in performance was nevertheless observed when expanding the model beyond 12 qubits or when adding additional quantum classifier blocks. Scalability limitations and practical challenges need still to be addressed, such as the barren plateau phenomenon and the overparameterization problem within quantum circuits. 
\item Grouping counterparts into sets: this step aims to aggregate the counterparts into a limited number of grades according to the score assigned by the model, to provide immediate information about creditworthiness. In this phase, different groupings (rating scales) are tested to find the most appropriate one.
 To the best of our knowledge, no quantum strategy has been defined for approaching this step. Therefore, we choose to devote our efforts to this task, to assess the potential of quantum computing.
This problem can be approached in two ways: as a clustering or an optimization task. When framed as a clustering task, quantum clustering methods, such as quantum k-means, can potentially handle large datasets more efficiently than classical techniques and identify better-defined clusters, thereby supporting the construction of more accurate and granular rating scales. For instance, Lloyd’s quantum algorithm \cite{Lloyd_Mohseni_Rebentrost_2013} estimates the distance to cluster centroids in $O(M \log(MN))$ time, whereas classical algorithms typically require $O(M^2 N)$, where $M$ denotes the number of data points and $N$ the dimensionality of the feature space. Improved scaling can be achieved by applying techniques similar to those used in quantum nearest-neighbor algorithms \cite{Wiebe_Kapoor_Svore_2014}. Moreover, the Q-means algorithm \cite{Kerenidis_Landman_Luongo_Prakash_2018} achieves a running time that depends polylogarithmically on the number of data points.
Several approaches have also been proposed in the NISQ setting, including quantum k-means variants \cite{Khan_Awan_Vall-Llosera_2019, PhysRevA.101.012326} and hybrid quantum–classical methods \cite{POGGIALI2024114466}. 

However, formulating the rating scale definition problem as a clustering task requires encoding financial constraints into suitable distance metrics, effectively shifting much of the problem complexity to the metric design stage.
For this reason, we instead focus on a quantum optimization approach. The rating scale definition problem naturally maps to a combinatorial optimization formulation; however, the large number of grades and counterparts encountered in real-world applications prevents a direct and tractable formulation using classical exact methods.
\end{enumerate}

\subsection{Quantum Optimization}
Quantum optimization (QO) has emerged as a notable branch of quantum computing~\cite{abbas}, addressing computationally intractable problems characterized by exponential complexity. In particular, many CO problems can be reformulated as QUBO problems~\cite{qubo} or equivalently mapped to the Ising model~\cite{Lucas_2014}; their resolution in the quantum domain relies on encoding the associated classical cost function into a quantum Hamiltonian, which describes the energy of a physical system and how it evolves over time. The field of QO is fundamentally built upon the ability to identify the ground state, or the global minimum energy configuration, of the system's Hamiltonian.

Within the current landscape of quantum computing, two primary paradigms have emerged to address this challenge, distinguished by their operational logic and hardware requirements. The first is the analog approach, represented by adiabatic quantum computing and its heuristic implementation, quantum annealing (QA). This method relies on the continuous physical evolution of a quantum system. The second is the digital-variational approach, which utilizes gate-based universal quantum computers to execute algorithms such as the Quantum Approximate Optimization Algorithm (QAOA)~\cite{qaoa} or the Variational Quantum Eigensolver (VQE)~\cite{vqe}. These paradigms differ in execution: one is a continuous physical process, while the other is a discrete sequence of logic gates. However, they are unified by their reliance on QUBO as the standard mathematical interface. This allows complex combinatorial constraints to be translated into a format compatible with quantum mechanical operators.

In the current NISQ era, gate-based universal quantum computers primarily address optimization through hybrid quantum-classical algorithms, such as QAOA and VQE. These approaches treat the quantum processor as a co-processor, where a parameterized quantum circuit (\textit{ansatz}) is used to prepare a trial state. The quality of this state is evaluated by measuring the expectation value of the problem Hamiltonian, and the resulting data is fed into a classical optimizer to iteratively update the parameters of the circuits. The performance of these algorithms is intrinsically tied to the circuit: increasing the depth improves solution accuracy; however, it leads to significant noise accumulation and decoherence. Furthermore, the classical optimization step faces challenges such as barren plateaus~\cite{Wang_2021}, hindering the convergence toward the optimal QUBO solution.
Extensive studies have explored the resolution of QUBO problems on gate-based quantum hardware~\cite{qaoa,khu,chie,zhou}, a research effort that has been further extended to the financial sector~\cite{vqeinnan2025quantum,QUBOfinance}.

In contrast to the discrete gate-based paradigm, quantum annealing provides a continuous analog alternative grounded in the adiabatic theorem. The primary operational advantage of QA lies in the mechanism of quantum tunneling. Unlike classical heuristic solvers, such as simulated annealing, which rely on thermal energy to overcome potential barriers in the energy landscape, QA allows the system to \say{tunnel} through high and narrow barriers. This capability is particularly effective for navigating the complex, non-convex landscapes typical of NP-hard problems encoded in QUBO or Ising formats. Given its ability to bypass local optima through tunneling and its superior scalability for binary optimization, quantum annealing remains a prominent quantum technology for a wide range of combinatorial applications~\cite{hauke,yarkoni,abbas,rajak,das,far2001,quantum6020018}.

Despite the theoretical foundations of quantum combinatorial optimization, the transition to large-scale applications is currently constrained by the noise and limited qubit counts of NISQ hardware, which impact the performance on complex, high-dimensional problem instances. Therefore, the primary contribution of contemporary research must focus on the methodological validation of the problem encoding itself. Developing precise QUBO formulations is a critical intermediate step; it allows researchers to verify the correctness of the energy landscape and the embedding strategies on small-scale prototypes. This validates the logical encoding of the solution, ensuring that the transition to future fault-tolerant quantum computers will require only scaling the hardware rather than redesigning the mathematical models.
\section{A QUBO formulation for defining the rating scale}
\label{section:formulation}

\subsection{QUBO models}
\label{QUBOModels}

A QUBO problem is the unconstrained optimization of a multivariate second-degree binary polynomial called the \textit{objective function}. Optimization refers to the search for instances of the variable $\vec{x}$ that maximize (or minimize) the objective function
\begin{equation} \label{qubo}
	\Xi(\vec{x})= \sum_{l_{1}l_{2}} Q_{l_{1}l_{2}} x_{l_{1}}x_{l_{2}} +q
\end{equation}
where $Q_{l_{1}l_{2}}$ is a two-index real parameter, $q$ is a real parameter, and each $x_{l}$ (element of the $\vec{x}$ vector) belongs to $\mathbb{B}=\{0,1\}$, the space of binary variables.
The function $\Xi$ can be written using the matrix multiplication rule
$\Xi(\vec{x}) = \vec{x}^{T} Q \vec{x} + q$
where $Q$ is the square matrix with components $Q_{l_{1}l_{2}}$.

If a QUBO model is a minimization problem its objective function is called cost function. We consider a quadratic constrained binary minimization problem having a cost function $\Xi(\vec{x})$: $\min_{\vec{x}}\Xi(\vec{x})$.
By adding to $\Xi$ suitable terms called penalties, we can construct $\Xi'$, transforming the original constrained problem into an unconstrained one.
\textit{Penalties} are functions that penalize the values of $\vec{x}$ that violate the constraints of the problem while favoring solutions that satisfy them. Each penalty is weighted by a suitable coefficient that reduces or amplifies its value.
It is possible to find some rules in the literature to convert constraints into penalties, as discussed in~\cite{qubo,unbal}.
Some constraints are linear binary equalities, such as $\sum_l p_l x_l = G$, some others are
linear binary inequalities, such as $\sum_l q_l x_l \le D$, where $p_l$, $q_l$, $G$ and $D$ are integers. In addition, certain constraints do not fall into either of these categories, and the corresponding penalty terms are therefore defined on a case-by-case basis.

Translating an equality constraint into its QUBO penalty is straightforward:
\begin{equation}\label{gen_eq_constr}
    \sum_{l}p_{l}x_{l} = G \mapsto \mu_{1} \cdot \bigg( \sum_{l}p_{l}x_{l} - G \bigg)^2
\end{equation}
where $\mu_1$ is a positive scalar coefficient.
In contrast, translating inequality constraints is more complex: two main methods have been proposed for this conversion. One approach is the unbalance penalization presented in~\cite{unbal}, which can be compared with the slack variable method described in~\cite{qubo}. The former approach requires two coefficients per inequality, whereas the latter method requires only one. To avoid expanding the coefficient space, this work adopts the slack variable approach.
Specifically, we treat the inequalities as follows:
\begin{equation}\label{gen_ineq_constr}
    \sum_{l}q_{l}x_{l} \le D \mapsto \mu_2 \cdot \bigg(D - \sum_{l}q_{l}x_{l} - S\bigg)^2
\end{equation}
where $S$ is an integer variable known as the \textit{slack variable}, such that 
\begin{equation}
0 \le S \le D - \min_{\vec{x}} \sum_{l}q_{l}x_{l}
\end{equation}
and $\mu_2$ is a positive scalar coefficient.
$S$ is decomposed into slack binary variables $s_{l}$ as follows:
\begin{equation}\label{s_integer}
    S = \sum_{l=0}^{\overline{N}_{S}-1}2^{l}s_{l}
\end{equation}
where the number of slack binary variables is
\begin{equation} \label{p8}
    \overline{N}_{S} = \left \lfloor 1+\log_{2} \biggl( D- \min_{\vec{x}}  \sum_{l}q_{l}x_{l}  \biggr) \right \rfloor .
\end{equation}

A new function $\Xi'$ incorporates the constraints as penalty terms. This is
\begin{align} \label{complete_qubo}
    \Xi'(\vec{x},\vec{s}) = & \sum_{l_{1} l_{2}}Q_{l_{1}l_{2}}x_{l_{1}} x_{l_{2}} \nonumber \\
    &+ \mu_{1} \cdot \bigg( \sum_{l}p_{l}x_{l} - G \bigg)^2 \\
    &+ \mu_{2} \cdot \bigg( D - \sum_{l}q_{l}x_{l} - \sum_{l=0}^{\overline{N}_{S}-1}2^{l}s_{l} \bigg)^2 \nonumber .
\end{align}
It is worth noting that not all values of the coefficients allow for equivalence between the constrained and unconstrained problems. Several studies~\cite{qubo,Ayodele_2022,VERMA2022100594} have attempted to address this issue by exploring methods for effectively fixing these parameters. As described in Section~\ref{sec:solvers}, this manuscript adopts an empirical approach to determine such multiplicative factors.

\subsection{Constraints for the rating scale definition problem}
As previously defined, a rating scale aims to cluster the counterparts while preserving their ordering by score.
Therefore, the problem consists of partitioning the ordered sequence of counterparts, $a_{1},a_{2},\dots,a_{n}$, into a given number of nonempty sets, called grades, $C_{1},C_{2},\dots,C_{m}$, according to certain constraints. At first glance, the integers $n$ and $m$ satisfy $2 \le m \le n$.
Each counterpart $a_i$ has two key attributes. The first attribute is the numerical score that provides a synthetic measure of the counterpart's creditworthiness. The second is the default variable $d_i$, the binary parameter that indicates the default status of the $i$-th counterpart: $d_i=1$ if and only if the default event occurred. The default rate (DR) of the $j$-th grade, denoted by $\ell_{j}$, is the ratio of the number of defaults in the grade to the number of counterparts belonging to the same grade. 

In accordance with~\cite{gim,eba}, we considered the following financial constraints:
\begin{itemize}
    \item[1.] monotonicity: the DR of the grades increases as the grade index increases;
	\item[2.] heterogeneity: from a statistical standpoint, the number of defaults in two consecutive grades is \textit{distant}. By fixing a lower threshold, this distance is evaluated through the t-test variable that quantifies the separation between averages of the two sets of default values;
	\item[3.] concentration: a high concentration of counterparts in one or more grades is not permitted, usually an acceptability threshold is defined but in general configuration with a low concentration are preferred;
	\item[4.] grade cardinality thresholds: the number of counterparts per grade is limited both above and below;
    \item[5.] grade homogeneity: for each grade, the DRs of two randomly selected complementary subsets are statistically \textit{close} to each other. This closeness is quantified in a similar manner with respect to the heterogeneity, employing a z-test with a predefined upper threshold.
\end{itemize}

In this work, we formalized a QUBO minimization problem tailored to 1, 3, and 4, since the binary encoding of the heterogeneity and homogeneity constraints requires a large number of additional binary variables (see the supplementary materials for their formal definitions and details). Moreover, heterogeneity and homogeneity constraints rely on statistical variables that require a large sample size, and due to the limited dimension of the datasets employed, the introduction of these was not considered meaningful.

\subsection{From constraints to QUBO penalties}

Prior to discussing the QUBO formulation of the constraints, we introduce useful notation conventions. First of all, we set $[a,b]=\{c' \in \mathbb{N} \mid a\le c' \le b\}$ and $[c]=\{c' \in \mathbb{N}\mid 1 \le c' \le c\}$ where $a,b,c$ are natural numbers. Then, we set the pre-established range of $i,i_1,i_2$ equal to $[n]$ and the pre-established range of $j$ equal to $[m]$.
The universal quantifier specification is omitted, unless the range at hand differs from the pre-established range of the variable; the same convention holds for variable
range specification in summations.

Given an arrangement of ordered counterparts $a_{i}$ in the grade $C_{j}$, we define the two-index variable:
\begin{equation} \label{eq_x2}
	x_{ij} = 
	\left\{
	\begin{array}{ll}
		1 & \text{if $a_{i} \in C_{j}$}\\
		0 & \text{otherwise}.\\
	\end{array} 
	\right.
\end{equation}
We can represent $\vec{x}$ in a matrix form of $n$ rows and $m$ columns as follows:
\begin{equation} \label{00}
  x = \left( {\begin{array}{ccccc} 
    1 & 0 & \cdots & \cdots & 0 \\
    \vdots & \vdots & \ddots & \ddots&0\\
    1 & 0 & \cdots & \cdots & 0\\
    0 & 1 & 0 & \cdots & 0\\
    \vdots & \vdots & \vdots & \ddots&0\\
    0 & 1 & 0 & \cdots &0\\
    &&\vdots&&\\
    0 & \cdots & \cdots & 0 & 1\\
    \vdots & \ddots & \ddots & \vdots &\vdots\\
    0 & \cdots & \cdots & 0 &1\\
  \end{array} } \right).
\end{equation}
This binary staircase matrix (BSM) identifies a valid partition of the counterparts. The nature of the problem is linked to imposing this type of structure on the binary variable $\vec{x}$. Therefore, we add this \textit{logical constraint}, or the 0-th constraint, to the previously described financial constraints to define the QUBO formulation of our problem.
Note that the $i$-th row of $x$ specifies the grade $j$ in which the counterpart $a_{i}$ is placed. By analyzing the \text{$j$-th} column, one can derive the cardinality of the \text{$j$-th} grade
\begin{equation} \label{14}
	N_{j}(x) = \sum_i x_{ij}.
\end{equation}
The DR of the $j$-th grade is
\begin{equation} \label{l}
	\ell_{j}(x) = \frac{1}{N_{j}(x)} \sum_{i \in C_{j}(x)} d_{i}
\end{equation}
where $C_{j}(x) = \{ i  \mid x_{ij}=1 \}$.

In light of this, the QUBO formulation of our problem employs a cost function equal to the sums of the penalties related to the constraints $0, 1, 3$ and $4$:
\begin{equation}\label{qubo_costf}
    \Xi(\vec{x},\vec{s}) = \Xi^{(0)}(\vec{x}) + \Xi^{(1)}(\vec{x}) + \Xi^{(3)}(\vec{x}) + \Xi^{(4)}(\vec{x},\vec{s})
\end{equation}
where indices in round brackets refer to the aforementioned enumeration of the constraints, and $\Xi^{(4)}(\vec{x},\vec{s})$ depends also on $\vec{s}$ which is the vector of the binary slack variables introduced to treat the inequalities related to the grade cardinality thresholds.
The encoding of each penalty is presented in the following subsections, whereas the supplementary materials give detailed explanations regarding their calculations.

\subsubsection{Logical constraint}

The penalty that encodes a BSM is composed of several terms. First of all, we know that each counterpart can only be assigned to one grade (grade uniqueness), which introduces the constraint $\sum_{j} x_{ij}= 1$ and, consequently, the penalty:
\begin{equation} \label{eq_1class_1grade}
	\mu_{01} \sum_i \biggl(\sum_j x_{ij}-1\biggr)^{2}.
\end{equation}

Since the counterparts are ordered, it follows that the first counterpart must belong to the first grade, meaning that $x_{11}=1$. Similarly, the last counterpart must belong to the last grade, meaning that $x_{nm}=1$. The second penalty is:
\begin{equation}\label{eq_1stcl_lastcl}
    \mu_{02}\cdot(1-x_{11})+ \mu_{02}\cdot(1-x_{nm}).
\end{equation}

Then, we impose that each column contains only one contiguous sequence of $1s$. To achieve this, we maximize the frequency of the cases $x_{ij}x_{i+1j}=1$. The corresponding QUBO contribution is 
\begin{equation}\label{eq_logic_ones}
	\mu_{03}\sum_{i \neq n} \sum_{j} (-1) x_{ij}x_{i+1j}.
\end{equation}

Finally, to ensure that the last counterpart of a grade is followed by a counterpart belonging to the next grade, we maximize the frequency of the cases
$x_{ij}x_{i+1j+1}=1$. This means that we add the penalty:
\begin{equation}\label{eq_logic_changeclass}
	\mu_{04}\sum_{i \neq n} \sum_{j \neq m} (-1) x_{ij}x_{i+1j+1}.
\end{equation}
The logical constraint penalty $\Xi^{(0)}(\vec{x})$  of the cost function is the sum of terms~(\ref{eq_1class_1grade},\ref{eq_1stcl_lastcl},\ref{eq_logic_ones})
and~(\ref{eq_logic_changeclass}).

\subsubsection{Monotonicity of the default rates}
\label{sec-monoton}

The formalization of the monotonicity constraint is:
\begin{equation} \label{eq_monoton}
	\quad \ell_{j}(x) \leq \ell_{j+1}(x)
\end{equation}
for every $j\neq m$, where
\begin{equation}
	\ell_{j}(x) = \frac{1}{N_{j}(x)} \sum_{i} d_{i}x_{ij}.
\end{equation}

Therefore, given that empty grades are excluded, the constraint can be expressed as a system of $m-1$ quadratic binary inequalities
\begin{equation}\label{eq-monoton-real}
	\sum_{i_{1}i_{2}}d_{i_{1}i_{2}}x_{i_{1}j} x_{i_{2}j+1} \leq 0
\end{equation}
for every $j\neq m$, where the antisymmetric parameter $d_{i_1 i_2}$ is equal to \text{$d_{i_1} - d_{i_2}$}.
Since these inequalities are quadratic, using the slack variable method introduces quartic terms that are not directly encodable in a QUBO formulation. The same consideration holds for the unbalanced penalization method.
To solve this issue, we can linearize the monomials in (\ref{eq-monoton-real}) introducing the additional binary variables $y$ as
\begin{equation}
	y_{t(i_1 i_2 j)} \equiv x_{i_{1}j} x_{i_{2}j+1} \ \text{for every} \ (i_1 i_2 j) \in Q_3
\end{equation}
where $Q_3 = \{ (i_1 i_2 j) \ \mid  (i_1 i_2) \in Q_2,  j\neq m \}$, \text{$Q_2 = \{(i_1 i_2)  \mid d_{i_1 i_2} \neq 0\}$}, $t$ represents a bijection defined in $Q_3$ onto the set \text{$[0,2(m-1)(n-d)d-1]$}, and $d$ is the number of defaults in the problem. In this case, the monotonicity constraint introduces
\text{$2(m-1)(n-d)d+(m-1) \left \lfloor 1+ \log_2 [(n-d)d] \right \rfloor$}
additional binary variables. This significantly increases the complexity of the problem and consequently leads to longer execution times.

To avoid introducing additional variables, we relax the original condition by adding to the cost function a term proportional to:
\begin{equation}
	\sum_{i_1 i_2 j \neq m} d_{i_{1}i_{2}} x_{i_1 j} x_{i_2 j+1}.
\end{equation}
In other terms, instead of searching for the $x$ that causes all the $m-1$ left hand sides of (\ref{eq-monoton-real}) to be non-positive, we require that the sum of the same $m-1$ terms appears as an addendum of the cost function.
As a consequence, the monotonicity penalty is approximated and does not introduce any additional variables. We obtain the penalty:
\begin{equation}\label{eq_monoton_pen}
    \Xi^{(1)} (\vec{x})= \mu_1 
	\sum_{i_1 i_2 j \neq m} d_{i_{1}i_{2}} x_{i_1 j} x_{i_2 j+1}.
\end{equation}
Since a minimizer of the QUBO cost function built with (\ref{eq_monoton_pen}) is not necessarily a configuration that fulfills (\ref{eq_monoton}), and \textit{vice versa}, we performed the analysis reported in Section~\ref{sec:approx-monoton} to validate this approximation.

\subsubsection{Concentration}

The Herfindahl index of a configuration is given by \text{$H(x) = \sum_{j} f_{j}^{2}(x)$}, where $f_{j}(x) = N_{j}(x)/n$ is the fraction of counterparts belonging to the $C_{j}$ in the configuration $x$. If all grades in a configuration $x$ have the same cardinality, then $f_{j}(x)=H(x)=1/m$ (note that these types of arrangements may not exist).
The concentration of counterparts within a grade is evaluated by \say{adjusting} the Herfindahl index, thus obtaining the adjusted Herfindahl index, which ranges from $0$ to $1$:
\begin{equation} \label{herf}
	H_{\text{adj}}(x) = \frac{H(x) - \frac{1}{m}}{1-\frac{1}{m}}.
\end{equation}
The adjusted Herfindahl index is the normalized measure of the nonuniformity of a distribution of the counterparts in the grades.
Since we are interested in configurations with low $H_{\text{adj}}$ values, we add it to the cost function to favor configurations that minimize it. By choosing a proper coefficient $\mu$, among the solutions that meet all other constraints, we select those with the lowest $H_{\text{adj}}$.
Since the $H_{\text{adj}}$ is a nonnegative variable, its penalty as a weak constraint is written as the product of the positive coefficient $\mu_3$ and the $H_{\text{adj}}$ with no square:
\begin{equation} \label{xi3}
	\Xi^{(3)}(\vec{x}) = \mu_3 \cdot \biggl[  \frac{m}{(m-1)n^2} \sum_{i_1 i_2 j} x_{i_1j}x_{i_2j} - \frac{1}{m-1}  \biggr].
\end{equation}

\subsubsection{Grade cardinality thresholds}
The grade cardinality $N_{j}(x)$ is constrained to lie between $1\%$ and $15\%$ of the total number $n$ of counterparts: $ n/100 \le  N_{j}(x) \le 15n/100$.
Since $2 \le m \le n$, there exists an integer $N_{j}$ such that the constraint is fulfilled only for $n \ge 7$.
With $\lambda_1 =  \lfloor n/100 \rfloor  \ \text{and} \ \lambda_2 =  \lceil 15n/100 \rceil $, $2m$ binary inequalities emerge from this constraint:
\begin{align}
    \sum_{i}(-x_{ij}) &\le  - \lambda_1,\\
    \sum_{i}x_{ij}&\le \lambda_2.
\end{align}

Using the slack variables as described in section~\ref{QUBOModels}, we obtain from the first $m$ inequalities:
\begin{align} \label{lower_thr}
	\mu_{41} \cdot\biggl( &\lambda_1 ^2 m
    + \sum_{i_{1} i_{2} j} x_{i_1 j} x_{i_2 j}
    + \sum_{j} \sum_{l_1 l_2 =0}^{\overline{N}_1-1} 2^{l_1+l_2} (s_1) {}_{l_1 j} (s_1) {}_{l_2 j}
	\nonumber\\
    &+ \sum_{ij} (-2) \lambda_1 x_{ij}
    + \sum_{j}\sum_{l=0}^{\overline{N}_1-1} \lambda_1 2^{l+1} (s_1) _{lj} \\
    &+\sum_{ij}\sum_{ l=0}^{\overline{N}_1-1} (-2)2^l x_{ij} (s_1) _{lj} \biggr) \nonumber
\end{align}
where $\overline{N}_1 = \left \lfloor 1+ \log_2 (-\lambda_1 + n) \right \rfloor$
and $s_{1}$ is a $\overline{N}_1 \times m$ binary slack variable.

Analogously, from the last $m$ linear binary inequalities, we obtain:
\begin{align} \label{lupper_thr}
	\mu_{42} \cdot  \biggl(&\lambda_2 ^2 m
    + \sum_{i_{1} i_{2} j} x_{i_1 j} x_{i_2 j}
    + \sum_{j} \sum_{l_1 l_2 =0}^{\overline{N}_2-1} 2^{l_1+l_2} (s_2) {}_{l_1 j}  (s_2) {}_{l_2 j} \nonumber \\
	& + \sum_{ij} (-2) \lambda_2 x_{ij}
	+ \sum_{j}\sum_{l=0}^{\overline{N}_2 -1}  (-1)\lambda_22^{l+1} (s_2) {}_{lj} \\
    &+ \sum_{ij}\sum_{l=0}^{\overline{N}_2-1} 2^{l+1} x_{ij} (s_2) {}_{lj}  \biggr) \nonumber
\end{align}
where $\overline{N}_2 = \left \lfloor 1+ \log_2  \lambda_2 \right \rfloor$
and $s_{2}$ is a $\overline{N}_2 \times m$ binary slack variable.

The grade cardinality thresholds penalty $\Xi^{(4)}(\vec{x}, \vec{s})$ is the sum of the terms~(\ref{lower_thr}) and~(\ref{lupper_thr}).
Note that this constraint introduces $m\cdot(\overline{N}_1 + \overline{N}_2)$ slack binary variables.

\section{Code implementation and results}
\label{sec_results}

The scripts used to test the QUBO formulation are outlined in this section. The full implementation, including all scripts used for model validation and benchmarking, is available in the project's GitHub repository~\cite{repo}.

All experiments reported in this work were conducted on synthetic datasets. Unless otherwise stated, both the credit scores and the default events were generated according to realistic distributions.

\subsection{A constrained implementation}
\label{sec:classic-impl}
The first task of the implementation phase was to develop a constrained algorithm that explores the combinatorics of partitioning the $n$ counterparts into $m$ grades. This initial step provides a reference framework against which to assess computational performance. More specifically, this allows us to evaluate whether the QUBO formulation presented in Section~\ref{section:formulation} provides a computational advantage compared to the brute-force constrained approach.

To achieve this task, we developed a function that, for each one of the $\binom{n-1}{m-1}$ configurations, checks if the configuration fulfills every constraint. For every configuration, all constraints are checked in aforementioned order; if one constraint is not satisfied, the configuration is discarded, and the algorithm moves on
to the next candidate.

For these tests we used datasets with defaults ranging from $1$ to $3$ for small cases where the realistic $3\%$ of counterparts is less than $1$. Additionally, we allowed for adjacent grades with the same DR that, in realistic cases, correspond to a single grade. To avoid high concentration of counterparts in the grades, we impose an upper bound on the adjusted Herfindahl index: $H_{\text{adj}}(x)<0.05$.

\begin{figure}[ht]
    \centering
    \includegraphics[width=1\linewidth]{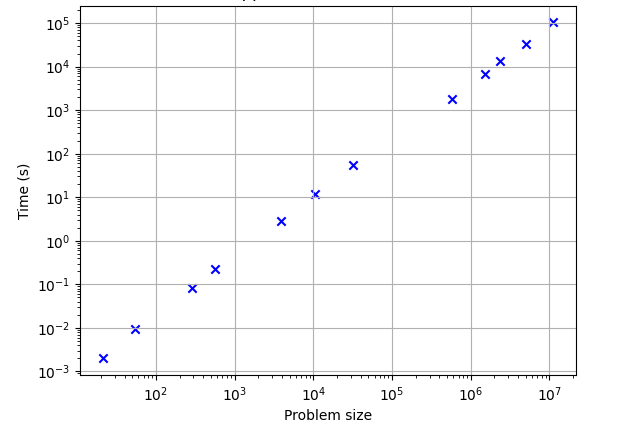}
    \caption{Computational performance of the constrained algorithm.}
    \label{fig:classic-plot}
\end{figure}

The results of the tests we performed on the LEONARDO supercomputer~\cite{leonardo2024} are reported in Table~\ref{tab:classic-result} and in the corresponding plot in Figure~\ref{fig:classic-plot}. The table shows the time required to find all the solutions that fulfill the constraints for different problem sizes. The time reported in the table is the average time obtained from three tests using different default vectors.

\begin{table}
\small

    \centering
    \begin{tabular}{rrrr}
        \hline
        \textbf{n} & \textbf{m} & \textbf{Configurations} & \textbf{Mean time (s)} \\
        \hline
        8  & 3 & 21       & 0.001989 \\
        12 & 3 & 55       & 0.009527 \\
        14 & 4 & 286      & 0.080983 \\
        17 & 4 & 560      & 0.225437 \\
        20 & 5 & 3876     & 2.872898 \\
        25 & 5 & 10626    & 11.909233 \\
        32 & 5 & 31465    & 55.039188 \\
        40 & 6 & 575757   & 1792.051127 \\
        48 & 6 & 1533939  & 6691.606785 \\
        52 & 6 & 2349060  & 13166.657564 \\
        60 & 6 & 5006386  & 33767.710212 \\
        70 & 6 & 11238513 & 103726.824730 \\
        \hline
    \end{tabular}
    \caption{Computational performance of the constrained algorithm. Time is computed as the mean over three different default vectors.}
    \label{tab:classic-result}
\end{table}

Based on the linear trend of the log–log plot in Figure~\ref{fig:classic-plot}, we are interested in the power-law relationship between the problem size (or the number of configurations to test) and the computation time. Accordingly, we searched for a function of the form $\text{time}=a \cdot (\text{problem size})^b$  where the coefficients $a$ and $b$ were determined via linear regression on the logarithms of the data. We found

\begin{equation}
\text{time} \approx 4.56 \cdot 10^{-6} \cdot \binom{n-1}{m-1}^{1.33}.
\end{equation}

Using this relationship, we can estimate the computation time for larger instances. For example, with $n=2000$ and $m=9$, the predicted computation time is approximately $4.64 \cdot 10^{19}$ days. Similarly, when $n=20000$ and $m=9$, the number of permutations to test is approximately $6.34 \cdot 10^{26}$, leading to a predicted computation time of about $1.99 \cdot 10^{30}$ days. These results prove that solving such large instances through a constrained approach is computationally infeasible.

\subsection{Building the QUBO formulation}

Alongside the constrained benchmark, we developed a second algorithm that addresses the rating scale definition problem using the previously defined QUBO formulation. The script is divided into two parts: building the cost function and searching for solutions to the QUBO problem.

The first part of the main algorithm requires two inputs: the dataset and the parameters $\mu$ associated with each problem constraint. In more detail, we compute a penalty for each constraint as outlined in the cost function~(\ref{qubo_costf}).
Each term consists of a constant and a matrix, both multiplied by the corresponding parameter $\mu$ selected for the problem instance. Only after scaling the penalties can the terms be summed to obtain the global cost function for that problem instance.

Note that these matrices are $nm \times nm$. The matrices related to the threshold constraints, however, contain additional elements because they also include the terms related to the slack variables. To allow the summation of all matrices, the smaller ones are padded with zeros in the extra rows and columns to match the dimensions of the largest matrix.

\subsubsection{Assessing the monotonicity approximation}
\label{sec:approx-monoton}

In Section~\ref{sec-monoton}, we described the penalty used for the monotonicity constraint. This term, as defined in~(\ref{eq_monoton_pen}), is an approximation of the constraint defined in~(\ref{eq-monoton-real}). As previously discussed, we adopted this approach to avoid a substantial increase in the number of variables.

To validate the approximated term, we conducted a series of tests aimed at constructing confusion matrices. Our goal was to verify that the solutions obtained with the approximation comply with the exact constraint. Specifically, for different datasets\footnote{This analysis was only feasible for small problem instances because the number of binary configurations to test increases exponentially with the problem size.} we generated all binary strings that satisfied the logical constraint (\textit{actual positives}). We first verified which of these configurations satisfied the exact monotonicity constraint. Then, we computed the quantity $\Xi^{(0)}(\vec{x}) + \Xi^{(1)}(\vec{x})$, in which the monotonicity term is constructed using the approximated penalty, searching for the string(s) that minimize it. For building the confusion matrix, we considered only the configuration(s) that reached the minimum as \textit{predicted positives}.

\begin{figure}[t]
\centering

\begin{subfigure}{0.8\linewidth}
\centering
\includegraphics[width=\linewidth]{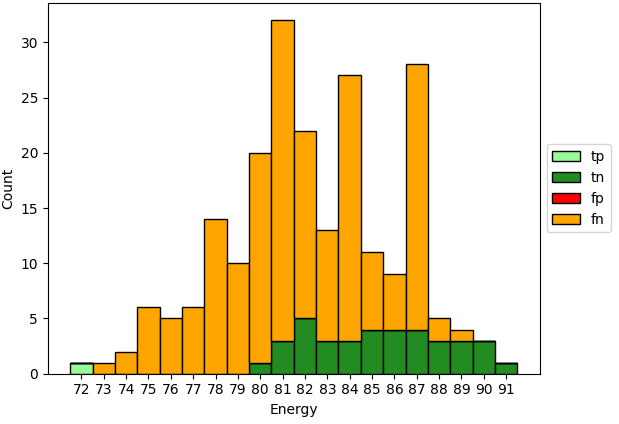}
\caption{Costs computed for all values of $\vec{x}$ of the cost function $\Xi(\vec{x},\vec{s})$, constructed using a dataset with 13 counterparts, 4 rating grades, and default events occurring at positions 9, 12, and 13.}
\end{subfigure}

\vspace{0.8em}

\begin{subfigure}{0.8\linewidth}
\centering
\includegraphics[width=\linewidth]{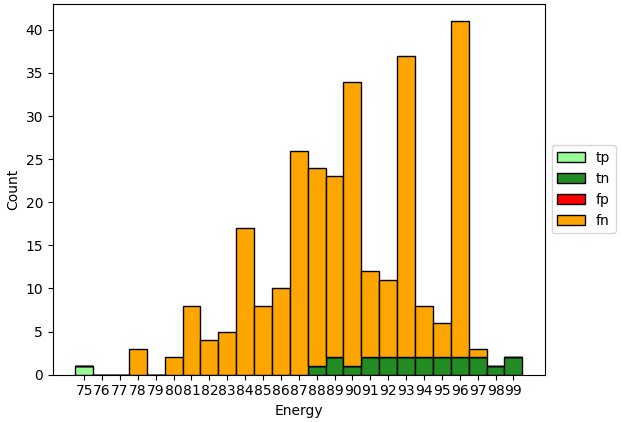}
\caption{Costs computed for all values of $\vec{x}$ of the cost function $\Xi(\vec{x},\vec{s})$, constructed using a dataset with 14 counterparts, 4 rating grades, and default events occurring at positions 11, 13, and 14.}
\end{subfigure}

\caption{Histograms of the cost function values associated with the confusion matrices reported in Tables~\ref{tab:confusion_matrices}. The x-axis shows the cost function values computed for all binary strings. Bars are colored according to the corresponding confusion matrix category: true positives (TP), false positives (FP), true negatives (TN), and false negatives (FN).}
\label{fig:monotonicity_histograms}

\end{figure}

An example of the results obtained on two different datasets is shown in Table~\ref{tab:confusion_matrices} and Figure~\ref{fig:monotonicity_histograms}\footnote{In this study, we focused only on realistic default distributions, excluding, for example, those that do not admit a valid solution.}. The approximation yields a single solution, and one true positive is always identified. Furthermore, no \textit{false positives} are observed. Additionally, in the histograms (Figure~\ref{fig:monotonicity_histograms}), it can be observed that the distribution of false negatives (configurations respecting the original monotonicity constraint) spans covering low energies, while true negatives can only be found at higher energies. This suggests that the penalty is discouraging true negatives, driving the global cost function $\Xi(\vec{x},\vec{s})$ towards a configuration that satisfies the monotonicity constraint.

\begin{table}[t]
\small

\begin{subtable}{\linewidth}
\centering
\begin{tabular}{c|cc}
\hline
 & Predicted Negative & Predicted Positive \\
\hline
Actual Negative & 43 & 0 \\
Actual Positive & 176 & 1 \\
\hline
\end{tabular}
\caption{Dataset with 13 counterparts, 4 grades, defaults at positions 10, 11, and 13.}
\end{subtable}

\vspace{0.5em}

\begin{subtable}{\linewidth}
\centering
\begin{tabular}{c|cc}
\hline
 & Predicted Negative & Predicted Positive \\
\hline
Actual Negative & 48 & 0 \\
Actual Positive & 237 & 1 \\
\hline
\end{tabular}
\caption{Dataset with 14 counterparts, 4 grades, defaults at positions 11, 13, and 14.}
\end{subtable}

\caption{Confusion matrices for comparing the exact and the approximated penalties for the monotonicity constraint.}
\label{tab:confusion_matrices}

\end{table}

In light of these considerations, we deemed this approximation to be acceptable. Future work could explore alternative strategies or investigate the direct implementation of the exact constraint.

\subsection{QUBO solvers}
\label{sec:solvers}

Once the QUBO formulation is defined, the search for solutions consists of finding the concatenation of the binary vectors $\vec{x}$ and $\vec{s}$ that minimizes the cost function. We evaluated different solvers, including implementations from various Python libraries, to identify the approach that best suits our problem in terms of accuracy and computational efficiency.

\subsubsection{Brute-force QUBO approach}
The simplest approach, albeit highly inefficient, is brute-force. If the solution vector has dimension $mn + m\cdot(\overline{N}_1+\overline{N}_2)$, then all $2^{mn + m\cdot(\overline{N}_1+\overline{N}_2)}$ possible binary strings must be considered. For each string, the cost function $\Xi(\vec{x}) = \vec{x}^{T} Q \vec{x} + q$ is computed, and the string(s) yielding the minimum cost are selected. 

The D-Wave \texttt{dimod} library~\cite{dimod} provides the \texttt{ExactSolver} function, which uses the brute-force approach to retrieve all solutions with minimum cost. While this functionality is useful for testing and validating small QUBO instances, it becomes impractical for larger problems\footnote{The documentation advises against using this solver when the number of variables exceeds 18.}. In our tests on the LEONARDO supercomputer, the algorithm failed for instances with more than 24 variables (e.g., 8 counterparts and 3 grades without the threshold constraint) because it required several terabytes of memory, which is infeasible in practice.

Additionally, the solver requires just under 2 seconds to find the solutions for the problem with 24 variables. Considering such a small instance, this represents a notably higher computation time compared to the constrained approach presented in the previous section.

Since the brute-force method does not exploit the computational advantages of the QUBO formulation, we evaluated alternative solvers.

\subsubsection{Gurobi}
\label{sec:gurobi}
Gurobi~\cite{gurobi} is a commercial optimization solver\footnote{Gurobi is a commercial software that is not freely available for general use. For this reason, we could not run Gurobi-based experiments on the LEONARDO supercomputer. Instead, we used a university server on which the academic license provided by the University of Ferrara is installed.}, which is used for combinatorial problems. In this work, we employed its \texttt{solve\_qubo} function from the \texttt{gurobi\_optimods} library. The function takes the QUBO matrix $Q$ and a time limit as input. Then, it transforms the QUBO formulation into a quadratic problem and applies heuristic methods, such as branch-and-bound, to efficiently explore the solution space. The function returns one solution that minimizes the cost function and its corresponding objective values (or the best solution found before the time limit expires).

The function provided by the \texttt{gurobi\_optimods} library is not the only Gurobi tool available for solving QUBO problems. The \texttt{gurobipy} library offers more advanced APIs that allow for more detailed studies, such as identifying all solutions with the same minimum objective value. Moreover, these APIs allow specifying a tolerance threshold, enabling the recognition of near-optimal solutions. The main drawback of this method is that it requires more computational time to obtain a valid solution than the first method (from minutes to hours for small instances). Consequently, while \texttt{gurobipy} is particularly useful for validation, it is impractical for medium-sized instances.

To provide an indication of the performance of the \texttt{solve\_qubo} method, we ran the same tests reported in Table~\ref{tab:classic-result}, starting with 32 counterparts and 5 grades, and setting a time limit of 3 minutes. Within this time and with the parameters $\mu$ described later, all tests returned a feasible solution, though it may not correspond to the minimum cost. With the same time, the solver could also find solutions that satisfied all constraints for larger datasets, with up to 175 counterparts and 9 grades. As an example, Table~\ref{tab:gurobi-test1} shows the results of a test with 150 counterparts and 9 grades.

\begin{table}[htbp]
\small
\centering
\begin{tabular}{c c c c}
\hline
Grade ID & Cardinality & Defaults & Default rate \\
\hline
1 & 16 & 0 & 0.000000 \\
2 & 16 & 0 & 0.000000 \\
3 & 16 & 0 & 0.000000 \\
4 & 16 & 0 & 0.000000 \\
5 & 17 & 0 & 0.000000 \\
6 & 17 & 0 & 0.000000 \\
7 & 17 & 1 & 0.058824 \\
8 & 17 & 1 & 0.058824 \\
9 & 18 & 4 & 0.222222 \\
\hline
\end{tabular}
\caption{Test with 150 counterparts, 9 grades, default vector (115, 131, 133, 147, 149, 150) and the parameter $\mu$ set as reported in the first set of Table~\ref{tab:penalties}.}
\label{tab:gurobi-test1}
\end{table}

A critical aspect of these tests was the determination of the parameters $\mu$. As described in~\cite{VERMA2022100594}, finding such parameters is not easy since not all combinations yield valid results. Large values tend to overwhelm the original objective function, and small values lead to infeasible solutions. In addition, the cost function we developed in Section~\ref{section:formulation} has 8 parameters, meaning the number of combinations of $\mu$ we can test is essentially unbounded.

For these reasons, we employed an empirical method to adjust the parameters based on the selected dataset. The first set reported in Table~\ref{tab:penalties} yields valid solutions for the tests with the same dataset as the constrained formulation, thereby validating the QUBO formulation presented in this study. However, it should be noted that this combination of parameters may not yield valid solutions for other problem instances. Identifying a configuration that is effective for all instances is beyond the scope of this study.

\begin{table*}[ht]
    \small
    \centering
    \begin{tabular}{l l l l}
    \hline
    \textbf{Penalty} & \textbf{Parameters} & \textbf{Set 1} & \textbf{Set 2} \\
    \hline
    Grade uniqueness & $\mu_{01}$ & $(nm)^2$ & $4(nm)^2$ \\
    Start from the first grade, end in the last grade & $\mu_{02}$ & $5  n  m$ & $5 n  m$ \\
    Other Binary staircase matrix coefficients & $\mu_{03}, \mu_{04}$ & $40  n  m$ & $75  n  m$ \\
    Monotonicity & $\mu_{1}$ & $5  d$ & $12  d$ \\
    Concentration & $\mu_{3}$ & $10 (n/m)$ & $3(n/m)$ \\
    Grade cardinality thresholds & $\mu_{41},\mu_{42}$ & $5(n/m)$ & $\mu_3 / 2$ \\
    \hline
    \end{tabular}
    \caption{Two valid parameter sets for the QUBO formulation solved with Gurobi.}
    \label{tab:penalties}
\end{table*}

With the first set of hyperparameters, as shown in Table~\ref{tab:gurobi-test1}, the formulation tends to favor solutions where the grades have the same cardinality. Furthermore, when the default is $4\%$ of the counterparts, many grades exhibit a $0$ default rate because the number of defaulting counterparts is not high enough to allow for a positive DR for each grade. Therefore, we conducted an additional set of tests in which more emphasis was placed on monotonicity with respect to the concentration constraint by selecting a higher parameter $\mu$. In these tests, the number of defaults was set to $10\text{-}12\%$, and the number of grades was fixed at $4$. The $\mu$ set is reported in the last column of Table~\ref{tab:penalties}, and an example of the result is given in Table~\ref{tab:gurobi-test2}.

\begin{table}[htbp]
\small
\centering
\begin{tabular}{c c c c}
\hline
Grade ID & Cardinality & Defaults & Default rate \\
\hline
1 & 36 & 0 & 0.000000 \\
2 & 40 & 2 & 0.050000 \\
3 & 42 & 6 & 0.142857 \\
4 & 32 & 10 & 0.312500 \\
\hline
\end{tabular}
\caption{Test with 150 counterparts, 4 grades and default vector (56, 63, 91, 96, 104, 106, 107, 113, 119, 122, 126, 127, 129, 133, 135, 144, 146, 149). The parameters $\mu$ are the second set in Table~\ref{tab:penalties}.}
\label{tab:gurobi-test2}
\end{table}

\section{Conclusion and future work}
\label{sec_conclusion}

In this study, we focused on the definition of the rating scale, a key component of the credit scoring process. This stage involves grouping counterparts into rating grades according to a set of predefined constraints.

In Section~\ref{sec:classic-impl}, we presented a constrained combinatorial brute-force algorithm designed to address the rating scale definition problem. The results of the executions indicate that the process is computationally intensive due to its power-law scaling behavior.

In light of these findings, we investigated whether a quantum-inspired approach could offer advantages for this task. To this end, we assessed the Quadratic Unconstrained Binary Optimization (QUBO) model as a promising candidate. Therefore, in Section~\ref{section:formulation}, we introduced a formulation that explicitly includes the logical, monotonicity, concentration, and grade cardinality thresholds constraints.

The logical constraint, together with the concentration and the grade cardinality thresholds constraints were encoded into the cost function through a quadratic penalty. However, the latter required the introduction of slack variables, thereby increasing the size of the problem.

Exact quadratic penalties could not be applied to the monotonicity constraint since it is expressed through quadratic inequalities. Adding slack variables to convert these inequalities into equality constraints led to quartic terms. Although techniques exist to reduce quartic terms to quadratic ones, doing so requires introducing additional variables, which quickly makes the problem computationally intractable.
In the supplementary material, we evaluated the linearization of quadratic inequalities to reduce the number of variables required for the quadratization of quartic penalties. Despite this reduction, the approach still involves a large number of additional variables.

For this reason, we adopted an approximate formulation of the monotonicity constraint. The proposed approximation can be incorporated into the cost function as softer penalty compared to the logical constraint, which must be strictly enforced. This is achieved by assigning lower amplification factors $\mu$ to the monotonicity constraint than to the logical ones. With an appropriate choice of parameters, we verified that solutions that closely satisfy the approximated constraint also fulfill the corresponding exact constraint and represent valid solutions to the original problem.

We will leave the investigation of more effective approximations for future work. However, it is worth noting that these strategies may be unnecessary once quantum technologies capable of solving QUBO problems with a very large number of variables become available. In this scenario, the exact quadratic formulation could be integrated directly into the cost function, which would only marginally impact computational performance.

For the same reason, the implementation of heterogeneity and homogeneity constraints is also left for future work.

We initially employed the D-Wave Hybrid solver in conjunction with a brute-force approach to validate the proposed formulation. This allowed us to verify each constraint individually, albeit only on small datasets. Then, we used the Gurobi \texttt{solve\_qubo} solver, a function of the commercial optimization software that efficiently addresses combinatorial optimization problems.

A central challenge in these experiments was selecting the parameters $\mu$, which determine the relative importance of each constraint. As discussed in Section~\ref{sec:gurobi}, values of $\mu$ that are too small may yield invalid solutions, whereas excessively large values may dominate the cost function and adversely affect the optimization process.

Despite these challenges, the experimental results show that the proposed QUBO formulation yields valid solutions when the parameters are chosen appropriately. The approach was tested on datasets comprising up to 175 counterparts and 4 to 9 grades. Evaluating larger datasets would require a more thorough analysis of the parameters $\mu$, as well as more time for computation to identify valid solutions.

The results presented in this work validate the effectiveness of the proposed approach and highlight the potential of quantum and quantum-inspired methods to support the definition process of rating scales. This formulation is particularly well-suited for solution via quantum annealing or QAOA, because of its direct correspondence with the Ising Hamiltonian. The results obtained with our QUBO formulation are promising and suggest that advanced quantum hardware could offer a substantial advantage in rating scale definition. Consequently, testing this approach on an actual quantum annealer is a goal for future work.
\section{Acknowledgements}

This work was supported by ICSC, \textit{Centro
Nazionale di Ricerca in HPC, Big Data and Quantum Computing}
(SPOKE 10) within the Italian “Piano Nazionale di Ripresa e Resilienza
(PNRR)”.

We would like to thank the University of Ferrara, and in particular the Department of Mathematics and Computer Science, for providing access to their servers with the academic license for Gurobi, which was essential for the computational experiments presented in this work.

\section{Declaration of competing interest}
The authors declare that they have no known competing
financial interests or personal relationships that could
have appeared to influence the work reported in this paper.

The views and opinions expressed are those of the authors
and do not necessarily reflect the views of Intesa Sanpaolo,
its affiliates or its employees.

\section{CRediT}
\textbf{Patrizio Spada:} Formal analysis, methodology, software, writing – original draft, review and editing;
\textbf{Laura Cappelli:} Investigation, methodology, resources, software, writing – original draft, review and editing;
\textbf{Francesca Cibrario:} Project administration, supervision, validation, formal analysis, writing – review and editing;
\textbf{Christian Mattia:} Supervision, validation, formal analysis, writing – review and editing;
\textbf{Daniele Magnaldi:} Validation, formal analysis, writing – review and editing;
\textbf{Matteo Argenton:} Supervision;
\textbf{Enrico Calore:} Supervision;
\textbf{Sebastiano Fabio Schifano:} Supervision, writing – review and editing;
\textbf{Concezio Bozzi:} Conceptualization, Supervision, writing – review and editing;
\textbf{Davide Corbelletto:} Supervision, writing – review and editing.

\bibliographystyle{elsarticle-num}
\biboptions{sort&compress}
\bibliography{reference}

\end{document}


\title{\textbf{Supplementary material of \say{A new approach to rating\\scale definition with quantum-inspired optimization}}}

\begin{frontmatter}





\end{frontmatter}

\appendix

\section*{Notations}
Prior to treat the calculations in details, we introduce the notations used
in this document. Commas that separate two indices are understood, even if
these indices are not subscripts.
Universal quantifier specification is always omitted and regarding summations we choose to not specify the ranges of $i$ and $j$ since they are easily deducible from the context.
The reader will notice that in some cases $i\in [n-1]$ and $j\in [m-1]$.
For all the other notations we refer to the articles.
\section{Binary staircase matrix}\label{constr0}
This section presents the main strategies used
for constructing a BSM of $n$ rows and $m$ columns as presented in the Section~\ref{section:formulation} of the article:
the so-called global approach, based on properties that characterize the matrix as a whole, and
the local approach. The latter has been implemented and tested but also
deemed as a second choice due to its higher number of parameters $\mu$
to be fixed before running.
\subsection{Global approach}\label{constr01}

We begin selecting a matrix $x$ such that $\sum_{j}  x_{ij}=1$.
This is imposed through
\begin{align}\label{3s}
	\mu_{01} \sum_{i} \biggl(\sum_{j}x_{ij}-1\biggr)^{2}.
\end{align}
The term in (\ref{3s}) expresses that \textit{each counterpart belongs to only one grade}.
Moreover we impose that \textit{each column contains only one contiguous sequence of 1s}.
To do this we maximize
\begin{align}\label{4s}
	\sum_{i} x_{ij}x_{i+1j}.
\end{align}
The QUBO contribution of (\ref{4s}) is
\begin{align}\label{5s}
	\mu_{02}\sum_{ij} (-1) x_{ij}x_{i+1j}.
\end{align}
Notice that
\begin{equation} \label{6s}
	x =
	\left( {\begin{array}{ccc}
			1 & 0 & 0\\
			1 & 0 & 0\\
			0 & 0 & 1 \\
			0 & 0 & 1 \\
			0 & 1 & 0
	\end{array} } \right)
\end{equation}
fulfills all the aforementioned constraints but it not a BSM. To avoid it
we maximize the case $x_{ij}x_{i+1j+1}=1$
\begin{align}\label{7s}
	\mu_{03}\sum_{ij} (-1) x_{ij}x_{i+1j+1}.
\end{align}
Doing this allows to express that \textit{the last counterpart of a grade is followed
by a counterpart belonging to the next grade}.
Due to the nature of the problem It follows that the
first counterpart must belong to the first grade, meaning that
$x_{11}=1$. Analogously, the last counterpart must belong to the last grade,
meaning that $x_{nm}=1$. Therefore, we add the penalty:
\begin{equation}\label{eq_1stcl_lastcl_1}
    \mu_{04}\cdot(1-x_{11}) + \mu_{04} \cdot (1-x_{nm}).
\end{equation}
Notice that condition (\ref{eq_1stcl_lastcl_1}) is  a consequence of
(\ref{3s},\ref{5s},\ref{7s}), if (\ref{3s},\ref{5s},\ref{7s})
are imposed with a descending order of priority. However imposing (\ref{eq_1stcl_lastcl_1})
as strong constraints helps reducing the variability
of the coefficients related to the other three logical conditions.
Finally we write 
\begin{align}\label{8s}
	\Xi^{(0)}(\vec{x}) =&
	\mu_{01} \sum_{i} \biggl(\sum_{j}x_{ij}-1\biggr)^{2}
	+\mu_{02}\sum_{ij} (-1) x_{ij}x_{i+1j}
	+\mu_{03}\sum_{ij} (-1) x_{ij}x_{i+1j+1}
    +\mu_{04}\cdot(1-x_{11}) + \mu_{04} \cdot (1-x_{nm}).
\end{align}
\subsection{Local approach}\label{constr03}

In the local approach the QUBO penalty is built as the sum of several terms similarly as in the global one.
First of all, we know that each counterpart can only be assigned to one grade, which introduces the penalty (\ref{3s}).
Since the first counterpart must belong to the first grade
and the last counterpart must belong to the last grade we
write $x_{11}=1$ and $x_{nm}=1$.
Therefore, the second penalty is:
\begin{equation}\label{eq_1stcl_lastcl_1_bis}
    \mu_{02}\cdot(1-x_{11}) + \mu_{02}\cdot(1-x_{nm}).
\end{equation}
Let us consider any $2 \times 2$ submatrix of a BSM:
$\begin{pmatrix}
x_{ij} & x_{ij+1} \\
x_{i+1j} & x_{i+1j+1}
\end{pmatrix}$.
Fixed $i$ and $j$, there are 16 possible sub-matrices.
The penalty (\ref{3s}) penalizes some of these. For
$
\begin{pmatrix}
1 & 0 \\
0 & 0
\end{pmatrix},
\begin{pmatrix}
0 & 0 \\
0 & 1
\end{pmatrix},
\begin{pmatrix}
0 & 1 \\
1 & 0
\end{pmatrix}
$
we have to introduce the following terms:
\begingroup
\allowdisplaybreaks
\begin{align}\label{eq_local_subm}
	&\mu_{04} \sum_{ij} [(1-x_{i+1j}-x_{i+1j+1})x_{ij}+x_{i+1j}x_{i+1j+1}]\nonumber\\
	&+\mu_{05} \sum_{ij} [(1-x_{ij}-x_{ij+1})x_{i+1j+1}+x_{ij}x_{ij+1}]\nonumber\\
	&+\mu_{06} \sum_{ij} x_{ij+1}x_{i+1j}    .
\end{align}
\endgroup
The sub-matrix
$\begin{pmatrix}
0 & 0 \\
1 & 0
\end{pmatrix}$
must be weakly penalized to  avoid the formation of a sequence/sequences
not-required sequence of 1s.
The relative penalty is:
\begin{equation}\label{eq_restart}
    \mu_{06} \sum_{ij} [(1-x_{ij}-x_{ij+1})x_{i+1j} + x_{ij}x_{ij+1}].
\end{equation}
Using the local approach, the term in the penalty that encodes the logic constraint is the sum of
(\ref{3s}, \ref{eq_1stcl_lastcl_1_bis}, \ref{eq_local_subm}) and (\ref{eq_restart}).

\section{Exact treatment of monotonicity}\label{constr1}
Let us begin by recalling that, as defined in the article, the DRs are given by
$\ell_{j}(x) = \frac{1}{N_{j}(x)} \sum_{i} d_{i}x_{ij}$
and the cardinality of the $j$-th grade is $N_{j} (x) = \sum_i x_{ij}$.
Therefore, besides $N_{j}(x)>0$, the monotonicity constraint ${\ell_{j}(x) \le \ell_{j+1}(x)}$ is
\begin{align}\label{2}
	\sum_{i_{1}i_{2}}d_{i_{1}i_{2}}x_{i_{1}j} x_{i_{2}j+1} \le 0
\end{align}
where $d_{i_1 i_2} \in \{-1,0,+1\}$ is $d_{i_1 i_2} = d_{i_1} - d_{i_2}$.
Such a parameter is antisymmetric: $d_{i_1 i_2} = -d_{i_2 i_1}$.
Let's consider
\begin{align}\label{4}
	C^{-} = \{(i_1 i_2) \in [n]^2 | d_{i_1 i_2} = -1\}, \quad \text{and} \quad
	C^{+} = \{(i_1 i_2) \in [n]^2 | d_{i_1 i_2} = 1\}.
\end{align}
We notice that
\begin{align}\label{6}
	\text{if} \ (i_1 i_2) \in C^{-} \ \text{then} \ (i_2 i_1) \in  C^{+}
\end{align}
and
\begin{align}\label{7}
	\text{if} \ (i_1 i_2) \in C^{+} \ \text{then} \ (i_2 i_1) \in  C^{-}.
\end{align}
Let us introduce $d \equiv|\vec{d}| = \sum_{i}d_i$.
For example if $\vec{d}=(0 \ 1 \ 1 \ 0 \ 0)$ we get $d=2$ with
\begin{equation} \label{9}
	d_{i_1 i_2} =
	\left( {\begin{array}{ccccc}
			0 & -1 & -1 & 0 & 0\\
			1 & 0 & 0 & 1 & 1\\
			1 & 0 & 0 & 1 & 1\\
			0 & -1 & -1 & 0 & 0\\
			0 & -1 & -1 & 0 & 0 
	\end{array} }  \right).
\end{equation}
The number of elements with $d_{i_1 i_2}=-1$ is $|C^{-}|=|C^{+}|=(n-d)d$, 6 in this example.

\subsection{Monotonicity inequalities}
Let's look at the $m-1$ inequalities in (\ref{2}) in the form:
\begin{align}\label{10}
	\sum_{(i_1 i_2)  \in C^-} (x_{i_2 j}x_{i_1 j+1}-x_{i_1 j}x_{i_2 j+1}) \le 0 .
\end{align}
These are equivalent to the system of $m-1$ equations
\begin{align}\label{21}
	\sum_{(i_1 i_2)  \in C^-} (-x_{i_2 j}x_{i_1 j+1}+x_{i_1 j}x_{i_2 j+1}) - S^{(2)}_{yj} = 0
\end{align}
where the integer variable $S^{(2)}_{yj}$ is defined by
\begin{align}\label{22}
	0 \le S^{(2)}_{yj} \le - \min \sum_{(i_1 i_2)  \in C^-} (x_{i_2 j}x_{i_1 j+1}-x_{i_1 j}x_{i_2 j+1}) = (n-d)d .
\end{align}
The (\ref{21}) becomes
\begin{align}\label{23}
	\sum_{(i_1 i_2)  \in C^-} x_{i_2 j}x_{i_1 j+1}- \sum_{(i_1 i_2)  \in C^-} x_{i_1 j}x_{i_2 j+1}+ \sum_{l=0}^{N_y-1} 2^l s^{(2)}_{ylj} = 0
\end{align}
where
\begin{align}\label{23_2}
	N^{(2)}_y = \left \lfloor 1+ \log_2 [(n-d)d] \right \rfloor.
\end{align}
The problem is well-defined only for $\vec{d}$ such that $d\in [1,n-1]$.
Notice that $s^{(2)}_{ylj}$ is a $N_y \times (m-1)$  binary matrix.
The (\ref{23}) is a system of $m-1$ quadratic equations.
Instead of quadratize the square of the LHS of (\ref{23}), we linearize it and square it.
For this reason we consider the products
\begin{align}\label{23_3}
	x_{i_{1}j} x_{i_{2}j+1} \ \text{for every} \ (i_1 i_2) \in C \ \text{and for every} \ j
\end{align}
where $C = C^- \cup C^+$. The cardinality of $C$ is $2(n-d)d$.
We set
\begin{align}\label{23_5}
	Q_3 = \{ (i_1 i_2 j) \in [n]^2 \times [m-1] |  (i_1 i_2) \in C \} .
\end{align}
Now we consider
\begin{align}\label{23_6}
	y_{t(i_1 i_2 j)} \equiv x_{i_{1}j} x_{i_{2}j+1} \ \text{for every} \ (i_1 i_2 j) \in Q_3
\end{align}
where $t$ is a bijection defined in the set of the triples $(i_1 i_2 j)\in Q_3$ onto
the 0-based set of indices having cardinality $2(m-1)(n-d)d$. We now introduce the couples of 0-based indices
\begin{gather}\label{23_7}
	(u_1 u_2)= ((i_1-1)m+j-1,(i_2-1)m+j-1) \ \text{for every}\  (i_1 i_2 j)\in Q_3
\end{gather}
and we define
\begin{align}\label{23_8}
	t(u_1  u_2) = t(i_1  i_2  j)
\end{align}
where $(u_1  u_2)$ and $(i_1  i_2  j)$ are linked by (\ref{23_7}). The couples $(u_1  u_2)$ are part of the set
\begin{align}\label{23_19}
	U_2 = \{(u_1 u_2)\in [0,nm-1]^2|&u_r = (i_r-1)m+j-1,\ \text{for every}\ r\in\{1,2\},\
	\nonumber\\&
	(i_1 i_2 j)\in Q_3  \} .
\end{align}
The binary constraint $y=x_1 x_2$ is implemented through the Rosenberg’s polynomial \cite{compr}
\begin{align}\label{11_6}
	\Xi^{(R)}(x_1, x_2, y) = \lambda_0 \cdot ( x_1 x_2 +3y -2x_1 y -2x_2 y )
\end{align}
where $\lambda_0$ is a suitable positive real number. Notice that $\Xi^{(R)}(x_1, x_2, y) > 0$ if and only if $y \ne x_1 x_2$; on the contrary $\Xi^{(R)}(x_1, x_2, y) = 0$ if and only if $y = x_1 x_2$. For every  $(i_1 i_2 j)\in Q_3$ we introduce the QUBO contribution
\begin{align}\label{12}
	\Xi^{(R)}(x_{i_1 j}, x_{i_2 j+1}, y_{t(i_1 i_2 j)}) = \lambda_0 \cdot ( x_{i_1 j} x_{i_2 j+1} +3y_{t(i_1 i_2 j)} -2x_{i_1} y_{t(i_1 i_2 j)} -2x_{i_2  j+1} y_{t(i_1 i_2 j)} )
\end{align}
that is equivalent to
\begin{align}\label{13_c}
	\Xi^{(R)}(x_{u_1}, x_{u_2+1}, y_{t(u_1 u_2)}) = \lambda_0 \cdot ( x_{u_1} x_{u_2 +1} +3y_{t(u_1 u_2)} -2x_{u_1} y_{t(u_1 u_2)} -2x_{u_2 +1} y_{t(u_1 u_2)} ).
\end{align}
The resulting contribution to the cost function for these terms  is
\begin{align} \label{13_d}
	\sum_{ (u_1 u_2) \in U_2 } \lambda_0 \cdot ( x_{u_1} x_{u_2+1} +3y_{t(u_1 u_2)} -2x_{u_1} y_{t(u_1 u_2)} -2x_{u_2 +1} y_{t(u_1 u_2)} ) .
\end{align}
Fixing a large positive real number $\lambda$,
from squaring (\ref{23}) and summing on $j$ we get
\begin{align}\label{24}
	\lambda\sum_j
	\bigg[&
	\sum_{((i_1 i_2),(i_3 i_4))  \in (C^{-})^2  } y_{t(i_2 i_1 j)} y_{t(i_4 i_3 j)}
	+\sum_{((i_1 i_2),(i_3 i_4))  \in (C^{-})^2  } y_{t(i_1 i_2 j)} y_{t(i_3 i_4 j)}
	+\sum_{l_1 l_2=0}^{N_y-1} 2^{l_1+l_2} s^{(2)}_{yl_1 j} s^{(2)}_{yl_2 j}\nonumber\\&
	-2\sum_{((i_1 i_2),(i_3 i_4))  \in (C^{-})^2  }y_{t(i_2 i_1 j)} y_{t(i_3 i_4 j)}
	+2\sum_{(i_1 i_2)  \in C^-} \sum_{l=0}^{N_y-1}  y_{t(i_2 i_1 j)}2^l  s^{(2)}_{ylj}\nonumber\\&
	-2\sum_{(i_1 i_2)  \in C^-} \sum_{l=0}^{N_y-1}  y_{t(i_1  i_2 j)}2^l s^{(2)}_{ylj}
	\bigg]
\end{align}
which is the QUBO addedum related to (\ref{2}).

\subsection{Change of summation variables}

Collecting (\ref{13_d}) and (\ref{24}):
\begin{align}\label{28}
	\Xi^{(1)}(x,y,s_y) =&
	\sum_{ (u_1 u_2) \in U_2}
	\lambda_0\cdot ( x_{u_1} x_{u_2 +1} +3y_{t(u_1 u_2)} -2x_{u_1} y_{t(u_1 u_2)} -2x_{u_2 +1} y_{t(u_1 u_2)} )
	\nonumber\\&
	+ \lambda    \sum_j \bigg[ 
	\sum_{((i_1 i_2),(i_3 i_4))  \in (C^{-})^2 } y_{t(i_2 i_1 j)} y_{t(i_4 i_3 j)}
	+\sum_{((i_1 i_2),(i_3 i_4))  \in (C^{-})^2 } y_{t(i_1 i_2 j)} y_{t(i_3 i_4 j)}\nonumber\\&
	+\sum_{l_1 l_2=0}^{N_y-1} 2^{l_1+l_2}s^{(2)}_{yl_1 j}s^{(2)}_{y l_2 j}
	-2\sum_{((i_1 i_2),(i_3 i_4))  \in (C^{-})^2  }y_{t(i_2 i_1 j)} y_{t(i_3 i_4 j)} \nonumber\\&
	+2\sum_{(i_1 i_2)  \in C^-} \sum_{l=0}^{N_y-1}  y_{t(i_2 i_1 j)}2^ls^{(2)}_{y lj}
	-2\sum_{(i_1 i_2)  \in C^-} \sum_{l=0}^{N_y-1}  y_{t(i_1  i_2 j)}2^ls^{(2)}_{ylj}
	\bigg] .
\end{align}
Let's look at the summations involving $(C^{-})^2$. By introducing
\begin{align}
	U_4 = \{ (u_1 u_2 u_3 u_4) \in [0,nm-1]^4|&u_r = (i_r-1)m+j-1, \text{for every}\nonumber\\&r\in\{1,2,3,4\}, (i_1,i_2)\in C^-, (i_3,i_4)\in C^- \}
\end{align}
we get
\begin{align}\label{29}
	\Xi^{(1)}(x,y,s_y) =&
	\sum_{ (u_1,u_2) \in U_2}
	\lambda_0 \cdot ( x_{u_1} x_{u_2 +1} +3y_{t(u_1 u_2)} -2x_{u_1} y_{t(u_1 u_2)} -2x_{u_2 +1} y_{t(u_1 u_2)} )
	\nonumber\\&
	+ \lambda \cdot \bigg[
	\sum_{(u_1 u_2 u_3 u_4)  \in U_4} y_{t(u_2 u_1)} y_{t(u_4 u_3)}
	+\sum_{(u_1 u_2 u_3 u_4)  \in U_4} y_{t(u_1 u_2)} y_{t(u_3 u_4)}\nonumber\\&
	+\sum_j\sum_{l_1 l_2=0}^{N_y-1} 2^{l_1+l_2}s^{(2)}_{yl_1 j}s^{(2)}_{y l_2 j}
	-2\sum_{(u_1 u_2 u_3 u_4)  \in U_4} y_{t(u_2 u_1)} y_{t(u_3 u_4)}\nonumber\\&
	+2\sum_j\sum_{(i_1 i_2)  \in C^-} \sum_{l=0}^{N_y-1}  y_{t(i_2 i_1 j)}2^ls^{(2)}_{y lj}
	-2\sum_j\sum_{(i_1 i_2)  \in C^-} \sum_{l=0}^{N_y-1}  y_{t(i_1   i_2 j)}2^ls^{(2)}_{ylj}
	\bigg] .
\end{align}
Since $s^{(2)}_{ylj}$ is a $N_y \times (m-1)$ binary matrix, for the quadratic summation in $s^{(2)}_{ylj}$ we define
\begin{align}
	V_2 =\{ (v_1   v_2) \in [0,N_y \cdot(m-1)-1] ^2|&v_r = l_r \cdot (m-1)+j-1 \ \text{for every}\
	l_r \in [0, N_y -1]   , \nonumber\\& r\in\{1,2\}    \},
\end{align}
and get
\begingroup
\allowdisplaybreaks
\begin{align}\label{30}
	\Xi^{(1)}(x,y,s_y) =&
	\sum_{ (u_1,u_2) \in U_2}
	\lambda_0 \cdot ( x_{u_1} x_{u_2 +1} +3y_{t(u_1 u_2)} -2x_{u_1} y_{t(u_1 u_2)} -2x_{u_2 +1} y_{t(u_1 u_2)} )
	\nonumber\\&
	+ \lambda \cdot \bigg[
	\sum_{(u_1 u_2 u_3 u_4)  \in U_4} y_{t(u_2 u_1)} y_{t(u_4 u_3)}
	+\sum_{(u_1 u_2 u_3 u_4)  \in U_4} y_{t(u_1 u_2)} y_{t(u_3 u_4)}\nonumber\\&
	+\sum_{(v_1 v_2) \in V_2} 2^{l(v_1)+l(v_2)}s^{(2)}_{yv_1}s^{(2)}_{y v_2}
	-2\sum_{(u_1 u_2 u_3 u_4)  \in U_4} y_{t(u_2 u_1)} y_{t(u_3 u_4)}\nonumber\\&
	+2\sum_j\sum_{(i_1 i_2)  \in C^-} \sum_{l=0}^{N_y-1}  y_{t(i_2 i_1 j)}2^ls^{(2)}_{y lj}
	-2\sum_j\sum_{(i_1 i_2)  \in C^-} \sum_{l=0}^{N_y-1}  y_{t(i_1  i_2 j)}2^ls^{(2)}_{ylj}
	\bigg] .
\end{align}
\endgroup
where
\begin{align}
	l(v) = \left \lfloor v/(m-1)\right\rfloor.
\end{align}
Now we treat the remaining two summations; by introducing
\begin{align}
	H = \{(u_1 u_2 v)  \in  [0,nm- 1]^2 \times [0,N_y \cdot (m-1)-1]
	|& u_r = (i_r-1)m+j-1,\nonumber\\&v=l\cdot (m-1)+j-1 \nonumber\\& \text{for every}\ r \in \{1,2\}, \ (i_1,i_2)\in C^- ,
	\nonumber\\&
	l \in [0, N_y -1]
	\}
\end{align}
we get
\begingroup
\allowdisplaybreaks
\begin{align}\label{31}
	\Xi^{(1)}(x,y,s_y) =&
	\sum_{ (u_1 u_2) \in U_2}
	\lambda_0 \cdot ( x_{u_1} x_{u_2 +1} +3y_{t(u_1 u_2)} -2x_{u_1} y_{t(u_1 u_2)} -2x_{u_2 +1 } y_{t(u_1 u_2)} )
	\nonumber\\&
	+ \lambda \cdot \bigg[
	\sum_{(u_1 u_2 u_3 u_4)  \in U_4} y_{t(u_2 u_1)} y_{t(u_4 u_3)}
	+\sum_{(u_1 u_2 u_3 u_4)  \in U_4} y_{t(u_1 u_2)} y_{t(u_3 u_4)}\nonumber\\&
	+\sum_{(v_1 v_2) \in V_2} 2^{l(v_1)+l(v_2)}s ^{(2)}_{yv_1}s ^{(2)}_{y v_2}
	-2\sum_{(u_1 u_2 u_3 u_4)  \in U_4} y_{t(u_2 u_1)} y_{t(u_3 u_4)}\nonumber\\&
	+2\sum_{(u_1 u_2 v) \in H }  y_{t(u_2 u_1)} 2^{l(v)}s ^{(2)}_{y v}
	-2\sum_{(u_1 u_2 v) \in H }  y_{t(u_1 u_2)} 2^{l(v)}s ^{(2)}_{y v}
	\bigg] .
\end{align}
\endgroup

\subsection{QUBO addendum}
We obtained
\begingroup
\allowdisplaybreaks
\begin{align}\label{31_3}
	\Xi^{(1)}(x,y,s_y) =&
	\sum_{ (u_1 u_2) \in U_2}
	(\lambda_0x_{u_1} x_{u_2 +1 } +3\lambda_0y_{t(u_1 u_2)} -2\lambda_0x_{u_1} y_{t(u_1 u_2)} -2\lambda_0x_{u_2 +1 } y_{t(u_1 u_2)})  
	\nonumber\\&
	+  
	\sum_{(u_1 u_2 u_3 u_4)  \in U_4} \lambda y_{t(u_2 u_1)} y_{t(u_4 u_3)}
	+\sum_{(u_1 u_2 u_3 u_4)  \in U_4} \lambda y_{t(u_1 u_2)} y_{t(u_3 u_4)}\nonumber\\&
	+\sum_{(v_1 v_2) \in V_2} 2^{l(v_1)+l(v_2)}\lambda s ^{(2)}_{yv_1}s ^{(2)}_{y v_2}
	+\sum_{(u_1 u_2 u_3 u_4)  \in U_4} (-2)\lambda y_{t(u_2 u_1)} y_{t(u_3 u_4)}\nonumber\\&
	+\sum_{(u_1 u_2 v) \in H }2^{l(v)+1}\lambda y_{t(u_2 u_1)}s ^{(2)}_{y v}
	+\sum_{(u_1 u_2 v) \in H }(-1) 2^{l(v)+1} \lambda y_{t(u_1 u_2)}s ^{(2)}_{y v}   
\end{align}
\endgroup
where $t$ is a bijection defined in
\begin{align}
	U_2 = \{(u_1 u_2)\in [0,nm - 1]^2|u_r = (i_r-1)m+j-1,\ \text{for every}\ r\in\{1,2\},\ (i_1 i_2 j)\in Q_3\}
\end{align}
onto the 0-based indices set having cardinality $2(m-1)(n-d)d$,
\begin{align}
	U_4 = \{ (u_1 u_2 u_3 u_4) \in [0,nm -1]^4|&u_r = (i_r-1)m+j-1 \ \text{for every}\nonumber\\&r\in\{1,2,3,4\}, (i_1,i_2)\in C^-, (i_3,i_4)\in C^-\},
\end{align}
\begin{align}
	V_2 =\{ (v_1   v_2) \in [0,N_y \cdot(m-1)-1] ^2|&v_r = l_r \cdot (m-1)+j-1 \ \text{for every}\
	l_r \in [0, N_y -1]   , \nonumber\\& r\in\{1,2\}    \},
\end{align}
\begin{align}
	H = \{(u_1 u_2 v)  \in  [0,nm  -1]^2 \times [0,N_y \cdot (m-1)-1]
	|& u_r = (i_r-1)m+j-1,\nonumber\\&v=l\cdot (m-1)+j-1 \nonumber\\& \text{for every}\ r \in \{1,2\}, \ (i_1,i_2)\in C^-  ,
	\nonumber\\&
	l  \in  [0, N_y -1]
	\},
\end{align}
\begin{align}
	l(v) = \left \lfloor v/(m-1) \right\rfloor,
\end{align}
\begin{align}
	C^{\pm} = \{(i_1 i_2) \in [1,n]^2 | d_{i_1 i_2} = \pm 1\}   ,
\end{align}
\begin{align}
	Q_3 = \{ (i_1 i_2 j) \in [1,n]^2 \times [1,m-1] |  (i_1 i_2) \in C \}  ,
\end{align}
\begin{align}
	N_y = \left \lfloor 1+ \log_2 [(n-d)d] \right \rfloor.
\end{align}

The binary vectors involved are:
\begin{align}
	x \in \mathbb{B}^{nm},\quad
	y \in \mathbb{B}^{2(m-1)(n-d)d} ,\quad
	s ^{(2)}_y \in \mathbb{B}  {}^{\wedge} \{ (m-1) \left \lfloor 1+ \log_2 [(n-d)d] \right \rfloor \}  .
\end{align}

For $n=5$, $m=3$, $d=2$
\begin{align}
	(n-d)d = 6
\end{align}
then the number of binary variables involved is:
\begin{align}
	nm + 2(m-1)(n-d)d+(m-1) \left \lfloor 1+ \log_2 [(n-d)d] \right \rfloor &= 15 + 24 + 6 \nonumber\\&
	=45.
\end{align}

\section{Concentration}\label{constr2}
This section shows the derivation of the penalty related to the weak concentration constraint.
Knowing that $N_{j}(x) = \sum_i x_{ij}$,
the adjusted Herfindahl index becomes:
\begin{align}
	H_{adj}(x)&=\frac{ \sum_{j}\frac{N^2_{j}(x)}{n^2} - \frac{1}{m}}{1-\frac{1}{m}}\nonumber\\
	&=\frac{ \frac{1}{n^2} \sum_{i_1 i_2 j} x_{i_1j}x_{i_2j} - \frac{1}{m}}{1-\frac{1}{m}}\nonumber\\
	&=\frac{ \frac{1}{n^2} \sum_{i_1 i_2 j} x_{i_1j}x_{i_2j} - \frac{1}{m}}{\frac{m-1}{m}}\nonumber\\
	&=m \frac{ \frac{1}{n^2} \sum_{i_1 i_2 j} x_{i_1j}x_{i_2j} - \frac{1}{m}}{m-1}\nonumber\\
	&=\frac{ \frac{m}{n^2} \sum_{i_1 i_2 j} x_{i_1j}x_{i_2j} -1}{m-1}\nonumber\\
	&=\frac{m}{(m-1)n^2} \sum_{i_1 i_2 j} x_{i_1j}x_{i_2j} - \frac{1}{m-1}.
\end{align}
All the other pertinent considerations are present in the article.

\section{Grade cardinality thresholds}\label{constr3}
This section shows the derivation of the penalties related to the cardinality threshold constraints.
Without losing generality we can fix two parameters $\lambda_1,\lambda_2\in[n]$
such that $\lambda_1 < \lambda_2$.
$\lambda_1$ and $\lambda_2$ are respectively the lower threshold and the upper threshold of the cardinalities.
We impose:
\begin{align}
	&\left\{
	\begin{aligned}
		&\lambda_1 \le \sum_{i} x_{ij}\\
		&\sum_{i} x_{ij} \le \lambda_2 .
	\end{aligned}
	\right.
\end{align}
The following treatment shows the derivations of the penalties related to
lower threshold constraint and the upper threshold one.
\subsection{Lower threshold}
The QUBO addendum relative to the constraint
$\lambda_1 \le \sum_{i} x_{ij}$
is proportional to the sum over $j$ of the squares of
\begin{align} \label{f1f2}
	\Phi_{jS}^{(41)} = -\lambda_1 + \sum_{i} x_{ij}- S_{j}^{(41)} .
\end{align}
where $S_{j}^{(41)} = \sum_{l=0}^{N_{S}^{(41)}-1} 2^l s_{lj}^{(41)}$ and $s_{lj}^{(41)}$ are
$mN_{S}^{(41)}$ binary variable with $ N_{S}^{(41)} = \left \lfloor \log_2 (-\lambda_1 + n
)+1 \right \rfloor $.
From (\ref{f1f2}) we obtain
\begin{align} \label{gct1}
	\Xi^{(41)} _S (\vec{x},\vec{s}_{41})
	=
	\mu_{41S} \cdot \biggl( & \lambda_1 ^2 m +
	\sum_{i_{1} i_{2} j} x_{i_1 j} x_{i_2 j} + \sum_{j} \sum_{l_1 l_2 =0}^{N_{S}^{(41)}-1} 2^{l_1} 2^{l_2} s_{l_1 j}^{(41)} s_{l_2 j}^{(41)}
	+\sum_{ij} (-2) \lambda_1 x_{ij}\nonumber\\
	&+ 
	\sum_{j}\sum_{l=0}^{N_{S}^{(41)}-1} \lambda_1 2^{l+1} s_{lj}^{(41)}+\sum_{ij}\sum_{ l=0}^{N_{S}^{(41)}-1} (-2)2^l x_{ij} s_{lj}^{(41)} \biggr).
\end{align}
The matrix $x$ is $n \times m$. The matrix $s$ is $N_S^{(1)}\times m$ elements.
If $n=4,m=3,\lambda_1=1$
\begin{align}
	N_{S}^{(41)} = \left \lfloor 1 + \log_2 3 \right \rfloor = 2.
\end{align}
The first summation is
\begin{align}
	\sum_{i_{1} i_{2} j} x_{i_1 j} x_{i_2 j} = \sum_{(u_1,u_2) \in U_2} x_{u_1} x_{u_2}
\end{align}
where $U_2 = \{ (u_1,u_2) \in [nm] ^2|
u_1 = (i_1-1)m+j , u_2 = (i_2-1)m+j\}$.
The second summation is
\begin{align}
	\sum_{j} \sum_{l_1 l_2 =0}^{N_{S}^{(41)}-1} 2^{l_1} 2^{l_2} s_{l_1 j}^{(41)} s_{l_2 j}^{(41)}.
\end{align}
We can introduce
\begin{align}
	v_1 = l_1 m + j\quad \text{e} \quad v_2 = l_2 m + j,
\end{align}
hence we can write the second summation as
\begin{align}
	\sum_{j} \sum_{l_1 l_2 =0}^{N_{S}^{(41)}-1} 2^{l_1} 2^{l_2} s_{l_1 j}^{(41)} s_{l_2 j}^{(41)}= \sum_{(v_1,v_2) \in V_2} 2^{\left \lfloor v_1 /m \right \rfloor + \left \lfloor v_2 /m \right \rfloor}
	s_{v_1}^{(41)} s_{v_2}^{(41)}
\end{align}
where $V_2 = \{ (v_1,v_2) \in[N_{S}^{(41)}m]^2|
v_1 = l_1 m+j , v_2 = l_2 m+j\}$.
The third summation is
\begin{align}
	\sum_{i j} (-2) \lambda_1 x_{i j} = \sum_{u \in U_1} (-2) \lambda_1 x_{u}
\end{align}
where $U_1= \{ u \in [nm]|u = (i-1)m+j\}$.
The forth summation is
\begin{align}
	\sum_{j}\sum_{l=0}^{N_{S}^{(41)}-1} \lambda_1 2^{l+1} s_{lj}^{(41)} = 
	\sum_{v \in V_1} \lambda_1 2^{\left \lfloor v /m \right \rfloor+1} s_{v}^{(41)}
\end{align}
where $V_1= \{ v \in [N_{S}^{(41)}m]|v = lm+j\}$.
The fifth summation is
\begin{align}
	\sum_{ij}\sum_{l=0}^{N_{S}^{(41)}-1} (-2)2^l x_{ij} s_{lj}^{(41)}
	= \sum_{ (w_1,w_2) \in W_2}(-1)2^{\left \lfloor w_2 /m \right \rfloor+1} x_{w_1} s_{w_2}^{(41)}.
\end{align}
where
\begin{align}
	W_2 = \{ (w_1,w_2) \in [nm] \times [N_{S}^{(41)}m]| w_1 = (i-1) +j, w_2 = lm+j \}.
\end{align}
The (\ref{gct1}) becomes
\begin{align}
	\Xi^{(41)} _S (\vec{x},\vec{s}_{41})
	=
	\mu_{41S} \cdot \biggl( &
	\lambda_1 ^2 m +
	\sum_{(u_1,u_2) \in U_2} x_{u_1} x_{u_2}
	+
	\sum_{(v_1,v_2) \in V_2} 2^{\left \lfloor v_1 /m \right \rfloor + \left \lfloor v_2 /m\right \rfloor}
	s_{v_1}^{(41)} s_{v_2}^{(41)}\nonumber\\
	&+
	\sum_{u \in U_1} (-2) \lambda_1 x_{u}+
	\sum_{v \in V_1} \lambda_1 2^{\left \lfloor v /m \right \rfloor+1} s_{v}^{(41)}\nonumber\\
	&+
	\sum_{ (w_1,w_2) \in W_2}(-1)2^{\left \lfloor w_2 /m \right \rfloor+1} x_{w_1} s_{w_2}^{(41)}
	\biggr)
\end{align}
The sets $U_2,V_2,U_1,V_1,W_2$ are $1$-based. Let's define
\begin{align}
	\overline{U}_2 &= \{ (u_1,u_2) \in [0,nm-1]^2|
	u_1 = (i_1-1)m+j-1, u_2 = (i_2-1)m+j-1\}, \nonumber\\
	\overline{V}_2 &= \{ (v_1,v_2) \in [0,N_{S}^{(41)}m-1]^2|
	v_1= l_1 m+j-1, v_2 = l_2 m+j-1\},\nonumber\\
	\overline{U}_1 &= \{ u \in  [0,nm-1]|u = (i-1)m+j-1\},\nonumber\\
	\overline{V}_1 &= \{ v \in [0,N_{S}^{(41)}m-1]|v = lm+j -1\},\nonumber\\
	\overline{W}_2 &= \{ (w_1,w_2) \in [0,nm-1] \times [0,N_{S}^{(41)}m-1]| w_1 = (i-1) +j-1, w_2 = lm+j-1\}.
\end{align}
These are their $0$-based counterparts.
Knowing that
\begin{align}
	\sum_{l=s}^t f_l = \sum_{l=s+k}^{t+k} f_{l-k},
\end{align}
in how case we obtain:
\begin{align}
	\sum_{l=1}^t f_l = \sum_{l=0}^{t-1} f_{l+1}.
\end{align}

The first summation becomes:
\begin{align}
	\sum_{(u_1,u_2) \in \overline{U}_2} x_{u_1} x_{u_2}
\end{align}
where $x_{u-1} \equiv x_{u}$.
The second summation becomes
\begin{align}
	\sum_{(v_1,v_2) \in \overline{V}_2} 2^{\left \lfloor (v_1 +1) /m \right \rfloor + \left \lfloor (v_2 +1) /m \right \rfloor}
	s_{v_1}^{(41)} s_{v_2}^{(41)}
\end{align}
where $s_{v-1}^{(41)} \equiv s_{v}^{(41)}$.
The third summation becomes
\begin{align}
	\sum_{u \in \overline{U}_1} (-2) \lambda_1 x_{u}.
\end{align}
The forth summation becomes
\begin{align}
	\sum_{v \in \overline{V}_1} \lambda_1 2^{\left \lfloor (v +1)/m \right \rfloor+1} s_{v}^{(41)}
\end{align}
and the fifth becomes
\begin{align}
	\sum_{ (w_1,w_2) \in \overline{W}_2}(-1)2^{\left \lfloor w_2 +1 /m \right \rfloor+1} x_{w_1} s_{w_2}^{(41)}.
\end{align}
Therefore the QUBO addendum relative to $\lambda_1 \le \sum_{i} x_{ij}$ is:
\begin{align} \label{gct4}
	\Xi ^{(41)} _S (\vec{x},\vec{s}_{41})
	=
	\mu_{41S} \cdot \biggl( &
	\lambda_1 ^2 m +
	\sum_{(u_1,u_2) \in \overline{U}_2} x_{u_1} x_{u_2}
	+
	\sum_{(v_1,v_2) \in \overline{V}_2} 2^{\left \lfloor (v_1 +1)/m \right \rfloor + \left \lfloor (v_2 +1)/m \right \rfloor}
	s_{v_1}^{(41)} s_{v_2}^{(41)}\nonumber\\
	&+
	\sum_{u \in \overline{U}_1} (-2) \lambda_1 x_{u}+
	\sum_{v \in \overline{V}_1} \lambda_1 2^{\left \lfloor (v +1)/m \right \rfloor+1} s_{v}^{(41)}
	\nonumber\\
	&+
	\sum_{ (w_1,w_2) \in \overline{W}_2}(-1)2^{\left \lfloor w_2 +1 /m \right \rfloor+1} x_{w_1} s_{w_2}^{(41)}
	\biggr).
\end{align}

\subsection{Upper threshold}

The QUBO addendum relative to $\sum_{i} x_{ij} \le \lambda_2$
is proportional to the sum over $j$ of the squares of
\begin{align}\label{phi42}
	\Phi_{jS}^{(42)} =\lambda_2-\sum_{i}x_{ij}- S_{j}^{(42)}.
\end{align}
where $S_{j}^{(42)} = \sum_{l=0}^{N_{S}^{(42)}-1} 2^l s_{lj}^{(42)}$
and $s_{lj}^{(42)}$ are $N_{S}^{(42)}m$ integer variables where
$N_{S}^{(42)} = \left \lfloor 1+ \log_2  \lambda_2	\right \rfloor$.
From (\ref{phi42}) we obtain:
\begin{align}
	\Xi^{(42)} _S (\vec{x},\vec{s}_{42})
	=
	\mu_{42S} \cdot \biggl( & \lambda_2 ^2 m +
	\sum_{i_{1} i_{2} j} x_{i_1 j} x_{i_2 j} + \sum_{j} \sum_{l_1 l_2 =0}^{N_{S}^{(42)}-1} 2^{l_1} 2^{l_2} s_{l_1 j}^{(42)} s_{l_2 j}^{(42)}
	+\sum_{ij} (-2) \lambda_2 x_{ij}\nonumber\\
	&+ 
	\sum_{j}\sum_{l=0}^{N_{S}^{(42)}-1} \lambda_2 (-1)2^{l+1} s_{lj}^{(42)}+\sum_{ij}\sum_{ l=0}^{N_{S}^{(42)}-1} 2^{l+1} x_{ij} s_{lj}^{(42)} \biggr) .
\end{align}
The matrix $s$ is composed by $N_S^{(2)}\times m$ elements.
For $n=4,m=3,\lambda_1=2$
\begin{align}
	N_{S}^{(42)} = \left \lfloor 1 + \log_2 2 \right \rfloor = 2.
\end{align}

The QUBO addendum relative to the upper threshold is analogue to (\ref{4}).
The sets $\overline{V}_2, \overline{V}_1, \overline{W}_2$
are to be replaced by
$\overline{\overline{V}}_2, \overline{\overline{V}}_1,
\overline{\overline{W}}_2$:
\begin{align}
	\overline{\overline{V}}_2&= \{ (v_1,v_2) \in [0,N_{S}^{(42)}m-1]\times [0,N_{S}^{(42)}m-1]|
	v_1= l_1 m+j-1, v_2 = l_2 m+j-1\},\nonumber\\
	\overline{\overline{V}}_1&= \{ v \in [0,N_{S}^{(42)}m-1]|v = lm+j -1\},\nonumber\\
	\overline{\overline{W}}_2&= \{ (w_1,w_2) \in [0,nm-1] \times [0,N_{S}^{(42)}m-1]| w_1 = (i-1) +j-1, w_2 = lm+j-1\}.
\end{align}
The others remain the same:
\begin{align}
	\Xi^{(42)} _S (\vec{x},\vec{s}_{42})
	=
	\mu_{42S} \cdot \biggl( &
	\lambda_2 ^2 m +
	\sum_{(u_1,u_2) \in \overline{U}_2} x_{u_1} x_{u_2}
	+
	\sum_{(v_1,v_2) \in \overline{\overline{V}}_2 }  2^{\left \lfloor (v_1 +1)/m \right \rfloor + \left \lfloor (v_2 +1)/m \right \rfloor}
	s_{v_1}^{(42)} s_{v_2}^{(42)}\nonumber\\
	&+
	\sum_{u \in \overline{U}_1} (-2) \lambda_1 x_{u}+
	\sum_{v \in \overline{\overline{V}}_1 }(-1)\lambda_1 2^{\left \lfloor (v +1)/m \right \rfloor+1} s_{v}^{(42)}
	\nonumber\\
	&+
	\sum_{ (w_1,w_2) \in \overline{\overline{W}}_2 } 2^{\left \lfloor w_2 +1 /m \right \rfloor+1} x_{w_1} s_{w_2}^{(42)}
	\biggr).
\end{align}

\section{Heterogeneity}\label{het}

This appendix and the following one rely on statistical notions \cite{casella, shi}.
In the credit scoring context, an adequate \textit{risk differentiation} among grades is
achieved through a \textit{meaningful differentiation} of their default values.
Such differentiation is called heterogeneity of default values; to check this requirement,
one compares the \textit{statistical distance} between the default values of two consecutive grades for each couple of grades.
For every couple of consecutive grades $(j,j+1)$ given a configuration $x$ we will impose
some constraints based on a t-test variable, its statistical validity and a fixed threshold value. 

If the standard deviations of the two populations of $j$ and $j+1$ in the configuration $x$ fulfill the conditions
\begin{equation}\label{std12}
	\frac{1}{2} <\frac{\sigma_{j}(x)}{\sigma_{j+1}(x)} < 2
\end{equation}
and
\begin{equation}
	N_{j}(x),\ N_{j+1}(x) \ge 30,
\end{equation}
where the binomial variances\footnote{We evaluate the binomial variance of the random variable per each grade;
this random variable corresponds to the occurrence/non-occurrence of the default and has a probability of success of a single trial
equal to the default rate of that grade.} per number of counterparts are
\begin{equation}
	\overline{\sigma_{j}^{2}}(x) =\ell_{j}(x) (1-\ell_{j}(x))
\end{equation}
then, in order to carry out the t-test, it is possible to write the t-test variable to our aim:
\begin{equation} \label{t}
	t_{j}(x) = \frac{\ell_{j}(x)-\ell_{j+1}(x)}{ (\sigma_{P})_{j}(x) \sqrt{ \frac{1}{N_{j}(x)} + \frac{1}{N_{j+1}(x)}  }}.
\end{equation}
The $(\sigma_{P})_{j}(x)$ is the pooled standard deviation:
\begin{equation} \label{pooled}
	(\sigma_{P})_{j}(x) = 
	\sqrt{\frac{ [N_{j}(x)-1] \sigma_{j}^{2}(x)  + [N_{j+1}(x)-1] \sigma_{j+1}^{2}(x)}{N_{j}(x)+N_{j+1}(x)-2}}
\end{equation}
where we omitted the overline signs over $\sigma_{j} ^2$ and $\sigma_{j+1}  ^2$ to simplify the notations.
Notice that (\ref{pooled}) is the square root of the weighted average of the average binomial variances of
grades $j$ and $j+1$ in configuration $x$.
The variable (\ref{t}) is approximately distributed according to the normal Gaussian
due to the high number of degrees of freedom.
We denote the cumulative distribution function of the normal Gaussian as $\mathcal{G}$.
The probability relative to the two-tails are 
\begin{align}
	P_{het}(t)&=\Pr(|T| \geq |t|)\nonumber\\
	&\simeq 2-2 \mathcal{G}(|t|).
\end{align}
To define the t-test constraint, let's fix the threshold at $\alpha=0.01$, i. e. 1\%:
\begin{itemize}
	\item if $P_{het} \le 0.01$ then the grade $(j,j+1)$ are heterogeneous;
	\item if $ 0.01< P_{het}$ then the grade $(j,j+1)$ are not heterogeneous.
\end{itemize}
The heterogeneity inequality relative to the $t$ variable holds if and only if:
\begin{align}\label{GG}
	|t| &\ge \mathcal{G}^{-1} \biggl(\frac{2 - \alpha}{2}\biggr).
\end{align}
If we set
\begin{equation}
	\gamma \equiv \mathcal{G}^{-1} \biggl(\frac{2 - \alpha}{2}\biggr),
\end{equation}
where $\gamma$ is a real positive number, we reach the system of inequalities:
\begin{align} \label{2_1}
	&\left\{
	\begin{aligned}
		&t_{j}^{2}(x) \ge \gamma^{2}\\
		& \frac{1}{2} <\frac{\sigma_{j}(x)}{\sigma_{j+1}(x)}<2\\
		& N_{j}(x) , N_{j+1}(x) \ge 30.\\
	\end{aligned}
	\right.
\end{align}
Notice that each grade must contain at least 30 counterparts, hence $n \ge 30 m$.

The QUBO implementation of the (\ref{2_1}) would imply the
transformation of the inequalities into equalities according to the
slack variable method. In particular $4m-3$ equations
depending on slack binary variables would result from this procedure.
It is important to notice that the nature of heterogeneity inequalities
is related to high degree polynomials to be linearized for deriving
their QUBO implementation. It is known that this kind of calculations
include the presence of high number of additional binary variables.

\section{Homogeneity} \label{hom}

Each grade must be representative of the default risk of the grade population itself. This is
ensured by imposing that the homogeneity constraint is satisfied. The notion of homogeneity is
characterized as follows: fixed a grade, for each pair of complementary sub-populations randomly
extracted 500 times from the considered grade, the z-test is performed on the DRs
to check the \textit{statistical closeness} of such DRs.
If the test is positive (probability of the z-variable greater than the 5\% threshold), then the two
samples are homogeneous. 
We introduce $q \in [500]$ as the index of the iteration.
Fixed a grade $j$, if the cardinalities of the two sub-populations fulfill the condition
\begin{equation} \label{cond}
	N'^{(q)}_{j},N''^{(q)}_{j} \ge 30 ,
\end{equation}
where
\begin{equation} \label{cond2}
	N'^{(q)}_{j}+N''^{(q)}_{j}=N_{j}
\end{equation}
we can introduce the following z-variable:
\begin{equation}
	z^{(q)}_{j} = \frac{\ell'^{(q)}_{j}-\ell''^{(q)}_{j}}{
		\sqrt{ \ell^{(q)}_{j}  (1-\ell^{(q)}_{j}) \bigg(\frac{1}{N'^{(q)}_{j}} + \frac{1}{N''^{(q)}_{j}} \bigg)  }}
\end{equation}
where $\ell'^{(q)}_{j}$ and $\ell''^{(q)}_{j}$ are the DRs of the considered sub-populations and
\begin{equation}
	\ell^{(q)}_{j} = \frac{d'^{(q)}_{j}+d''^{(q)}_{j}}{N'^{(q)}_{j}+N''^{(q)}_{j}} = \frac{d'^{(q)}_{j}+d''^{(q)}_{j}}{N_{j}}
\end{equation}
i.e., the ratio between the sum of the occurrences of the target event of the two sub-populations
and the sum of the cardinalities of the two sub-populations.
Finally, the homogeneity condition is given by
\begin{align} \label{5_1}
	&\left\{
	\begin{aligned}
		&\Pr(|T| \geq |z_{j}^{(q)}|)\ge 0.05\\
		&N_{j}^{(q)},N_{j+1}^{(q)} \ge 30. \\
	\end{aligned}
	\right.
\end{align}

Regarding homogeneity QUBO implementation, considerations
analogue to the case of heterogeneity need to be brought up.
In fact, the nature of these inequalities leads to high degree
polynomials that need to be tread through linearization techniques
with high cost in terms of additional binary variables.

\bibliographystyle{elsarticle-num} 
\bibliography{reference}

\biboptions{sort&compress}